\documentclass[longauth]{aa} 
\usepackage{txfonts} 
\usepackage[OT1]{fontenc}
\usepackage{graphicx}
\usepackage{epsfig}

\def\mathrelfun#1#2{\lower3.6pt\vbox{\baselineskip0pt\lineskip.9pt
  \ialign{$\mathsurround=0pt#1\hfil##\hfil$\crcr#2\crcr\sim\crcr}}}

\def\lesssim{\mathrel{\mathpalette\fun <}} 
\def\gtrsim{\mathrel{\mathpalette\fun >}}
\def\fun#1#2{\lower3.6pt\vbox{\baselineskip0pt\lineskip.9pt
  \ialign{$\mathsurround=0pt#1\hfil##\hfil$\crcr#2\crcr\sim\crcr}}}
\def\etal{{\it et~al. }}

\begin{document}
%
   \title{Archeops In-flight Performance, Data Processing and Map Making}

\author{ 
J.~F.~Mac\'{\i}as--P\'erez ~\inst{1} \and  
G.~Lagache~\inst{20} \and  
B.~Maffei~\inst{3} \and  
P.~Ade~\inst{3} \and  
A.~Amblard~\inst{4} \and  
R.~Ansari~\inst{5} \and  
{\'E}.~Aubourg~\inst{6, \, 7} \and
J.~Aumont~\inst{1} \and
S.~Bargot~\inst{5} \and
J.~Bartlett~\inst{7} \and  
A.~Beno\^{\i}t~\inst{8} \and  
J.--Ph.~Bernard~\inst{9} \and  
R.~Bhatia~\inst{25} \and
A.~Blanchard\inst{10} \and  
J.~J.~Bock~\inst{11, \,12} \and
A.~Boscaleri\inst{24} \and  
F.~R.~Bouchet~\inst{13} \and  
A.~Bourrachot~\inst{5} \and  
P.~Camus~\inst{8} \and  
J.-F.~Cardoso~\inst{14} \and
F.~Couchot~\inst{5} \and  
P.~de Bernardis~\inst{15} \and  
J.~Delabrouille~\inst{7} \and  
F.--X.~D\'esert~\inst{16} \and  
O.~Dor\'e~\inst{27, \, 20, \, 30} \and
M.~Douspis~\inst{20,\, 10} \and  
L.~Dumoulin~\inst{17} \and  
X.~Dupac~\inst{25} \and
Ph.~Filliatre~\inst{18} \and  
P.~Fosalba~\inst{19, \, 20} \and
K.~Ganga~\inst{7} \and
F.~Gannaway~\inst{3} \and
B.~Gautier~\inst{8} \and
M.~Giard~\inst{9} \and
Y.~Giraud--H\'eraud~\inst{7} \and  
R.~Gispert~\inst{20}$\dag$\thanks{Richard Gispert passed away few weeks
after his return from the early mission to Trapani} \and  
L.~Guglielmi~\inst{7} \and  
J.--Ch.~Hamilton~\inst{21} \and  
S.~Hanany~\inst{22} \and  
S.~Henrot--Versill\'e~\inst{5} \and  
V.~Hristov~\inst{11} \and
J.~Kaplan~\inst{7} \and  
J.-M.~Lamarre~\inst{23} \and
A.~E.~Lange~\inst{11} \and  
K.~Madet~\inst{8} \and  
Ch.~Magneville~\inst{6, \, 7} \and
D.~P.~Marrone~\inst{22} \and
S.~Masi~\inst{15} \and  
F.~Mayet~\inst{1} \and  
J.~A.~Murphy~\inst{26} \and
F.~Naraghi~\inst{1} \and
F.~Nati~\inst{15} \and  
G.~Patanchon~\inst{2} \and
O.~Perdereau~\inst{5} \and  
G.~Perrin~\inst{1} \and
S.~Plaszczynski~\inst{5} \and 
M.~Piat~\inst{7} \and  
N.~Ponthieu~\inst{20} \and  
S.~Prunet~\inst{13} \and 
J.--L.~Puget~\inst{20} \and 
C.~Renault~\inst{1} \and  
C.~Rosset~\inst{5} \and  
D.~Santos~\inst{1} \and 
A.~Starobinsky~\inst{28} \and
I.~Strukov\inst{29} \and
R.~V.~Sudiwala~\inst{3} \and
R.~Teyssier~\inst{13} \and
M.~Tristram~\inst{5,\, 1} \and  
C.~Tucker~\inst{3} \and 
J.--Ch.~Vanel~\inst{7} \and
D.~Vibert~\inst{13} \and 
E.~Wakui~\inst{3} \and 
D.~Yvon~\inst{6}
}

   \offprints{reprints@archeops.org}

\institute{
Laboratoire de Physique Subatomique et de Cosmologie, 53 Avenue des 
Martyrs, 38026 Grenoble Cedex, France
\and
Department of Physics \& Astronomy, University of British Columbia, Vancouver, Canada
\and
Cardiff University, Physics Department, PO Box 913, 5, The Parade,  
Cardiff, CF24 3YB, UK 
\and
University of California, Berkeley, Dept. of Astronomy, 601
Campbell Hall, Berkeley, CA 94720-3411, U.S.A.
\and
Laboratoire de l'Acc\'el\'erateur Lin\'eaire, BP 34, Campus
Orsay, 91898 Orsay Cedex, France
\and
CEA-CE Saclay, DAPNIA, Service de Physique des Particules,
Bat 141, F-91191 Gif sur Yvette Cedex, France
\and
APC, 11 pl. M. Berthelot, F-75231 Paris Cedex 5, France
\and
Centre de Recherche sur les Tr\`es Basses Temp\'eratures,
BP166, 38042 Grenoble Cedex 9, France
\and
Centre d'\'Etude Spatiale des Rayonnements,
BP 4346, 31028 Toulouse Cedex 4, France
\and
Laboratoire d'Astrophysique de Tarbes Toulouse,
14 Avenue E. Belin, 31400 Toulouse, France
\and
California Institute of Technology, 105-24 Caltech, 1201 East
California Blvd, Pasadena CA 91125, USA
\and
Jet Propulsion Laboratory, 4800 Oak Grove Drive, Pasadena,
California 91109, USA
\and
Institut d'Astrophysique de Paris, 98bis, Boulevard Arago, 75014 Paris,
France
\and
CNRS--ENST
46, rue Barrault, 75634 Paris, France
\and
Gruppo di Cosmologia Sperimentale, Dipart. di Fisica, Univ. 'La
Sapienza', P. A. Moro, 2, 00185 Roma, Italy
\and
Laboratoire d'Astrophysique, Obs. de Grenoble, BP 53,
38041 Grenoble Cedex 9, France
\and
CSNSM--IN2P3, Bât 108, 91405 Orsay Campus, France
\and
CEA-CE Saclay, DAPNIA, Service d'Astrophysique, Bat 709,
F-91191 Gif sur Yvette Cedex, France
\and
Instituto de Ciencias del Espacio (IEEC/CSIC), Facultad de Ciencias,
Campus UAB, E-08193 Cerdanyola, Spain
\and
Institut d'Astrophysique Spatiale, B\^at. 121, Universit\'e Paris
XI, 91405 Orsay Cedex, France
\and
LPNHE, Universit\'es Paris VI et Paris VII, 4 place
Jussieu, Tour 33, 75252 Paris Cedex 05, France
\and
School of Physics and Astronomy, 116 Church St. S.E., University of
Minnesota, Minneapolis MN 55455, USA
\and
LERMA, Observatoire de Paris, 61 Av. de l'Observatoire, 75014 Paris, France
\and
IROE-CNR, Firenze, Italy
\and
ESTEC, Noordwijk, The Netherlands
\and
Experimental Physics, National University of Ireland, Maynooth, Ireland
\and
Dpt of Astrophysical Sciences, Princeton University, Princeton, NJ08544 USA
\and
Landau Institute for Theoretical Physics, 119334 Moscow, Russia
\and
Space Research Institute, Profsoyuznaya St. 84/32, Moscow, Russia
\and
CITA, University of Toronto, 60 St George Street, Toronto, ON M5S  3H8, Canada
}

  \date{\today}

  \abstract{} {Archeops is a balloon--borne experiment widely inspired by the
    Planck satellite and by its High Frequency Instrument (HFI).  It is mainly
    dedicated to measure the Cosmic Microwave Background (CMB) temperature
    anisotropies at high angular resolution ($\sim 12$ arcminutes) over a
    large fraction of the sky (around 30~\%) in the millimetre and
    submillimetre range at 143, 217, 353 and 545~GHz.  Further, the Archeops
    353~GHz channel consists of three pairs of polarized sensitive bolometers
    designed to detect the polarized diffuse emission of Galactic dust.  }
  {We present in this paper the update of the instrumental setup as well as
    the inflight performance for the last Archeops flight campaign in February
    2002 from Kiruna (Sweden).  We also describe the processing and analysis
    of the Archeops time ordered data for that campaign which lead to the
    measurement of the CMB anisotropies power spectrum in the multipole range
    $\ell = 10-700$ (Beno{\^{\i}}t et al. 2003a, Tristram et al. 2005)
   and to the first measurement of the dust polarized emission at large
   angular scales and its polarized power spectra in the multipole range $\ell
   = 3-70$ (Beno{\^{\i}}t et al. 2004, Ponthieu et al. 2005).  }
  {
  We present maps of 30~\% of the sky of the Galactic emission, including the
  Galactic plane, in the four Archeops channels at 143, 217, 353 and 545~GHz
  and maps of the CMB anisotropies at 143 and 217~GHz.  These are the first
  ever available sub--degree resolution maps in the millimetre and
  submillimetre range of the large angular-scales Galactic dust diffuse
  emission and CMB temperature anisotropies respectively.}
{}

\maketitle


\section{Introduction}

The measurement of the Cosmic Microwave Background (CMB) anisotropies
in temperature and polarization is a fundamental proof of modern
cosmology and of the early Universe physics.  Since the first detection
of the CMB anisotropies by the COBE satellite in 1992 (\cite{smoot}), a
large number of ground--based and balloon--borne experiments such as DASI
(\cite{dasi}), CBI (\cite{cbi}), VSA (\cite{vsapaper}), BOOMERanG
(\cite{boomerang}) or Maxima (\cite{maxima}) have measured the CMB
angular power spectrum from a few-degrees down to sub-degree scales.
However, simultaneous observation of very large and small angular
scales have proved to be particularly difficult, as it requires both
large sky coverage and high angular resolution.  This has been
achieved, first, by Archeops (\cite{archpaper}, \cite{tristram_cl})
which has measured the CMB power spectrum in the multipole range $10 <
\ell < 700$.  Then, the WMAP satellite mission (\cite{wmap})
has detected the CMB anisotropies, both in temperature and
polarization.

Archeops, described in details in \cite{trapani}, is a balloon
borne--experiment designed as a prototype for the High Frequency Instrument
(HFI) of the Planck satellite.  Its telescope and focal plane optics are
widely inspired by the Planck design.  The implementation of the measurement
chains: cryogenics, optics, bolometers, readout electronics was a successful
validation of the innovative design.  Further the data processing was a
learning process for future members of the HFI team.  Archeops performs
circular scans on the sky with its optical axis tilted 41 degrees with respect
to the horizon by spinning the gondola at 2 rpm.  This scanning, that is
combined with the proper motion of the Earth, leads to ~30\% sky coverage in
about 12 hours of flight.  With a rotating gondola, the Sun above the horizon
produces a dominant parasitic signal.  The optimal way of avoiding this, while
having the longest integration time, is by having a long duration (Arctic)
night-time balloon flight.

The Archeops payload was successfully launched three times. First, from
Trapani (Italy) in July 1999 (\cite{trapani}) for a 4 hours test flight.  Then
from the Swedish Esrange station (near Kiruna at 68~deg. latitude North, just
above the Arctic circle) operated by the French Centre National d'Etudes
Spatiales (CNES) and the Swedish Space Corporation in January 2001 (hereafter
KS1 flight) for a 12 hours flight, and finally in February 2002 (hereafter KS3
flight) for a 24 hours flight from which 12 night-hours were exploited for
scientific purposes.  In the KS3 flight, a stratospheric altitude of 35 km was
reached, reducing significantly the contamination from atmospheric (mainly
ozone) emission with respect to ground--based measurements.  Additional
information about the Archeops flights may be found at our web--site\footnote{
  {\tt http://www.archeops.org }}.


The Archeops bolometers are grouped in four frequency bands at 143~GHz (8
bolometers), 217~GHz (8 bolometers and a blind one), 353~GHz (3 polarized
bolometer pairs), 545~GHz (1 bolometer). The 143 and 217~GHz channels are
dedicated to the measurement of the temperature angular power spectrum of the
CMB (\cite{archpaper,tristram_cl}). The 353 and 545~GHz channels allow the
monitoring of both atmospheric emission and Galactic thermal dust emission. In
addition, the polarization of the diffuse Galactic dust emission has been
measured for the first time using the 353~GHz polarized bolometers
(\cite{archpolar,ponthieu05}).

We present here the processing of the Archeops data for the KS3 flight going
from raw time ordered data to maps of the sky.  The Archeops data processing
was specifically designed to cope with the characteristics of the scanning
strategy and has similarities with Planck-HFI data processing.  Other
dedicated processing techniques are described in details in
\cite{maxima,maxiproc2,maxiproc3,maxiproc4,boomproc1,boomproc2,acbarproc} to
deal with the Maxima, BOOMERanG, ACBAR spider web bolometer-experiments data
and with WMAP HEMT all sky survey satellite data\cite{wmapproc}.

This paper is organized as follows. Sections~\ref{sec:archeopsexperiment} and
\ref{sec:inflightperformance} describe the instrumental set-up and the
in-flight performances of Archeops during the KS3 flight.
Section~\ref{sec:preprocessing} presents the preprocessing of the Archeops
data. In section~\ref{sec:pointing} we describe the offline pointing
reconstruction. Section~\ref{boloresponse} deals with the optical and time
responses of the instrument. Sections~\ref{sect:syste} and
\ref{sec:dataquality} present the characterization and treatment of
systematics and noise in the data.  In section~\ref{sec:calibration} we
discuss the intercalibration and absolute calibration of the Archeops data.
Finally, section~\ref{The Archeops maps} presents the construction of the
Galactic and CMB Archeops maps. We conclude in section~\ref{sec:conclusions}.

\section{Technical description of the experiment}
\label{sec:archeopsexperiment}

        \begin{figure}[t]
        \begin{center}
         \caption{Scheme of the Archeops gondola \label{fig:gondola}}
        \end{center}
        \end{figure}
In this section we describe the main aspects of the instrumental setup
of the Archeops experiment.  Particular interest is paid to changes
performed on this setup since the Trapani test flight (\cite{trapani}).

\subsection{Gondola}
The Archeops gondola was redesigned since the Trapani flight, in order
to gain some weight and try to reduce the main parasitic effect that
was observed then.  The most likely interpretation was indeed that
stray light reflecting from or emitted by inhomogeneities on the
balloon surface
was the main culprit for this large scale parasitic effect.  The new baffle
contains lightweight highly reflective material (Fig.~\ref{fig:gondola}) in a
staircase--like layout so that the entrance of the gondola is highly
``reflective'' for downward rays. To further reduce the systematics effects,
between flight KS1 and KS3, the engine driving the gondola spinning was moved
from the top of the gondola (and rigidly fixed to it) to an higher location
along the flight chain, 60 meter above the gondola. This allowed to strongly
reduce non stationary noise induced by the swivel engine.

\subsection{Attitude control system}
\label{sec:attitudecontrolsystem}
The Archeops attitude control systems are composed of gyroscopes, a GPS
and a Fast Stellar Sensor.  The gyroscopes are unchanged since Trapani
flight.  A high precision z--axis laser gyroscope based on the Sagnac
effect was added for long term relative azimuth reconstruction.  This
is needed for daylight data, when detected stars are not enough to
track rotation speed changes.  The GPS was changed since Trapani
because of a failure at high altitude.  A 1--m diameter circular loop
of copper wire was added and used as an Earth magnetic field detector
to perform a rough (5 degree accuracy) absolute azimuth reconstruction.
The electromagnetic influence of the pivot rotor and its associated
wire on the loop signal disappeared from the KS1 to the KS3 flight, as
the rotor was moved upwards along the flight chain.

The Fast Stellar Sensor (FSS) is a 40--cm optical telescope equipped with 46
photodiodes mounted on the bore--sight of the primary mirror
(Fig.~\ref{fig:gondola}) for a--posteriori accurate pointing reconstruction.
The photodiodes are aligned along a line which is perpendicular to the
scanning direction.  Each photodiode covers a 7.6 arcminutes (parallel to the
scanning direction - para scan) by 1.9 arcminutes (perpendicular to the
scanning direction - cross scan) area on the sky.  The FSS sweeps a 1.4 degree
wide ring at constant elevation during a payload revolution and its center is
mechanically within one degree of the main submillimetre telescope pointing
direction.  The FSS has been improved with respect to the Trapani
configuration (\cite{trapani}): a red filter has significantly diminished the
background and parasitic noises have been suppressed.  A full report on the
FSS is given by~\cite{Nati:2003}.  During the flights, about 100 to 200 stars
per revolution were detected by the FSS. A detailed description of the
pointing reconstruction can be found in Sect.~\ref{sec:pointing}.

\subsection{Detectors}

For the last KS3 flight campaign, the detection was ensured by an array
of 21 spider web bolometers (\cite{Bock:1996}) of the same type as for
the Maxima (100~mK) and BOOMERanG (300~mK) experiments.  For each
bolometer, a Neutron Transmutation Doped Germanium thermistor is fixed
on a silicon nitride micromesh designed to absorb submillimetre light.
The bolometers are cooled down below 100 mK by an $^3\rm He/^4\rm He$
dilution cryostat (\cite{cryostat}) and were optimized for the expected
background loads at this temperature, varying from 2 to 8~pW depending
on frequency.  The bolometers were built at JPL/Caltech in the context
of the development of the Planck HFI instrument (\cite{lamarreplanck}).

Bolometer characteristics were measured from standard I--V curves obtained
during ground--based calibrations at zero power load, after the bolometers
were made blind (see section~\ref{sss:bolmod}).  In order to prevent radio
frequencies contaminations, each bolometer is kept in a copper $\lambda/4$
cavity acting as a Faraday cage for maximal absorption.

The sensitivity of the bolometers at 100 mK is limited by the photon
noise and their short time response ranges from 5 to 14 ms,
which is adequate for the Archeops scanning and acquisition strategies.

\subsection{Optical configuration}

\begin{figure}[t]
  \begin{center}
    \caption{Optical layout of the Archeops focal plane.
      \label{opticallayout}}
  \end{center}
\end{figure}

The Archeops optical configuration consists of a 1.5 m off-axis
Gregorian telescope illuminating a set of back-to-back horns which are coupled
to each of the detectors in the focal plane.  The horns, which are
corrugated and flared, are cooled down to 10~K by helium vapors and
their wave guide sections act as frequency high-pass filters. Low-pass
filters are located at the back of the horns on the 1.6~K stage.  These
two sets of filters define the frequency band of operation of {\sc
Archeops} from 100 to 600~GHz.  The complete set of bolometer, filters
and horn constitute a photometric pixel.  Those are layed out on
constant elevation (scan) lines.  The telescope images these lines into
curved lines on the focal plane.  Figure~\ref{opticallayout} shows the
focal plane layout of the photometric pixels.  The layout of bolometers
at different frequencies was chosen so as to have redundancies on
different angular and time scales.  The main axis of each photometric
pixel is pointed at the image of the primary mirror through the
secondary mirror.  Entrances of the 10~K back--to--back horns are
located above the focal plane at various heights (typically about 6~mm), in
order to prevent optical cross--talks between channels.

\subsection{Observation frequency bands}
\begin{figure}
\begin{center}
\caption{Spectral transmission of the various types of photometric
  pixels. This was obtained by combining different measurements at
  the component level.
  \label{fig:bandpass}}
\end{center}
\end{figure}


The Archeops data are acquired at four frequency bands centered at 143
(8 photometers), 217 (8 photometers), 353 (6 photometers) and 545 (1
photometer)~GHz.  The first two were dedicated to the measurement of
the angular power spectrum of the CMB temperature anisotropies.  The
last two were designed to measure the dust diffuse Galactic emission.
Figure~\ref{fig:bandpass} shows the spectral transmission of the
various type of photometric pixels corresponding to the four Archeops
frequency bands.

The 353~GHz photometers are arranged in pairs coupled to the same horn
via an Ortho Mode Transducer (OMT, \cite{omt}) and are optimized to
measure the polarized sky signal as described in \cite{archpolar}.

\subsection{Cryostat monitoring}
Thermometers are used to monitor the cryostat temperature at each of the 
thermal stages
described above.  The thermometers, made of large thin films of NbSi,
are added to the 100~mK, 1.6 and 10~K stages.  They are described
in \cite{Camus:2000}.
The house keeping data obtained from these thermometers are essential
for the subtraction of low frequency drifts in the Archeops bolometer
data as described in Sect.~\ref{sect:syste}.

\section{In-flight performance}
\label{sec:inflightperformance}
        \begin{figure}[t]
        \begin{center}
         \caption{From top to bottom, evolution of the temperature 
           of the focal plane and of the 1.6~K and 10~K cryogenic stages
           during the KS3 flight}
        \end{center}
        \label{fig:cryotemp}
        \end{figure}

We discuss in this section the performance of the Archeops instrument
during the KS3 flight in terms of cryogenics and photometry.  The
flight took place on February 7$^{th}$ 2002 and lasted 21.5 hours,
starting at 12h44 UT time.  During 19 hours the balloon was at nominal
altitude, between 32.5 and 34.5 km above sea level, with a total of
12.5 hours of night data used for scientific purposes.

\subsection{Cryogenic performance}

The cryostat functioned autonomously during the entire flight duration.
The dilution flow was changed twice.  First it was decreased at the
beginning of the mission to increase the life time of the dilution.
Then, it was increased at sunrise to compensate for the extra thermal
power from the sun.

The full cryogenic system warmed up mechanically at launch, being at
nominal temperature at about 15h00 UT time. Fig.~\ref{fig:cryotemp}
shows the temperature of the focal plane, the 1.6~K and the 10~K
stages from top to bottom respectively.  At float altitude, the focal
plane cooled down staying below 100~mK during the entire flight.  A
plateau of about 90~mK was reached since 19h00 UT. The 1.6~K stage was
stable during the entire flight at a temperature of about 1.5~K. The
10~K stage remained at about 9~K until sunrise at
27h00~UT\footnote{Meaning 3h00 UT the next day}, and then increased up
to about 12~K. During the night flight the temperature of the
bolometer bath was stable at 90~mK.
 
\subsection{Bolometer signal and noise contributions}
 
\subsubsection{A simple photometric model}
\label{simplephotomodel}
In order to evaluate \emph{a priori} the total background incoming onto
each bolometer, we performed a component by component photometric
analysis by dealing with emission processes and various transmission
coefficients.  From the sky to the detector, it includes
\begin{itemize}
\item the CMB emission assuming a simple 2.725~K blackbody (\cite{Mather:1999})
\item the atmospheric emission for which the emissivity was computed for
  41~deg. elevation (airmass of 1.52) at 32~km altitude using the Pardo's
  atmospheric model (\cite{Pardo:2001}). A temperature of 250~K is assumed.
\item the radiation from the telescope which is assumed to have an emissivity
  of $0.00285\times\ 2\times\ \sqrt{1\,\mathrm{mm}/\lambda}$
  (\cite{bock1995}). The factor 2 is to account for the primary and secondary
  mirrors. A temperature of 250~K is assumed.
\item the emission of the polypropylene window, which allows the radiation to
  propagate to the cold optics while maintaining the vacuum, is neglected
  here.
\item the radiation from the 10~K stage which was found to emit in the
KS1 flight a detectable fraction of the background but which was negligible
for the KS3 flight.  We also account for a transmission factor across
the 10K stage estimated to be 0.6.
\item the transmission curves shown in Fig.~\ref{fig:bandpass}. They mostly
  represents the filtering done at the 1.6~K stage, although some
  filters were sometimes placed on the 10~K stage and a bandpass filter
  at the entrance of the 100~mK horn.
\item the bolometer with an assumed perfect absorption.
\end{itemize}

\subsubsection{A simple bolometer model}\label{sss:bolmod}
\label{simplebolomodel}
The theory of the thermodynamical and electrical behavior of bolometers
is described in details by~\cite{Mather:1984} and \cite{piat:2001}.  Here we
concentrate on the main equations to introduce the
parameterization given in subsequent tables.

We characterize the thermistor behavior of thermometers and bolometers using

\begin{equation}
\label{eq:resist}
R=R_{\infty }\exp \left( \frac{T_{r}}{T_{1}}\right) ^{\alpha }\ \mathrm{.}
\end{equation}
The electron--photon decoupling is neglected as well as electric field
effects in the bolometer.  The absolute temperature $T_{1}$ of the
thermistor is calibrated with carbon resistance and is valid within 3
milliKelvin.


The heat equilibrium for the bolometer reads:

\begin{equation}
\label{eq:conduct}
P_{C}(T_{1},T_{0})=g\left[ \left( \frac{T_{1}}{T_{100}}\right) ^{\beta
    }-\left( \frac{T_{0}}{T_{100}}\right) ^{\beta }\right]
=P_{J}+P_{R}\ \mathrm{,}
\end{equation}

 where \( T_{0} \) is the base plate temperature measured by the
standard thermometer, \( T_{1} \) is the bolometer temperature obtained
through \( R(T_{1}) \) (Eq.~\ref{eq:resist}), \( T_{100} \) is a reference
temperature (we take it as \( T_{100}=100\, \mathrm{mK}) \) so that
the constant \( g \) is in unit of pW, \( P_{C} \) is the cooling
power, \( P_{J}=UI \) is the Joule power dissipated in the thermistor
and \( P_{R} \) is the absorbed part of the incident radiative power.

\begin{table}[!ht]
  \center
\begin{tabular}{c|cc|cc|c}
\hline \hline
 Bolometer & $R_{\infty } (\Omega)$ & $T_{r} \mathrm{(K)}$ & 
$G \mathrm{(pW/K)}$ & $\beta$ & $C \mathrm{(pJ/K)}$ \\
\hline
    143K01 &     11.27 &     21.16 &     57.96 &      2.80 &      0.40 \\ 
    143K03 &     44.23 &     16.60 &     70.62 &      2.30 &      0.42 \\ 
    143K04 &     52.91 &     16.85 &     63.83 &      2.30 &      0.47 \\ 
    143K05 &     53.41 &     16.85 &     50.94 &      2.30 &      0.35 \\ 
    143K07 &     21.55 &     18.79 &     60.30 &      2.55 &      0.38 \\ 
    143T01 &     21.47 &     20.03 &    116.29 &      2.85 &      0.57 \\ 
\hline
    217K01 &    299.42 &     16.73 &     28.93 &      2.10 &      0.18 \\ 
    217K02 &    189.54 &     13.21 &     65.15 &      2.00 &      0.36 \\ 
    217K03 &    242.93 &     13.10 &     62.87 &      2.00 &      0.03 \\ 
    217K04 &    159.03 &     13.76 &     69.45 &      2.20 &      0.39 \\ 
    217K05 &   1172.80 &      9.72 &     59.31 &      1.65 &      0.34 \\ 
    217K06 &    120.92 &     14.18 &     65.11 &      2.10 &      0.46 \\ 
    217T04 &     52.38 &     14.79 &    161.00 &      2.30 &      1.06 \\ 
    217T06 &    136.69 &     13.67 &    182.16 &      2.30 &      0.00 \\ 
\hline
    353K01 &     99.45 &     14.39 &     90.50 &      2.20 &      0.21 \\ 
    353K02 &     94.59 &     15.01 &     98.47 &      2.20 &      0.12 \\ 
    353K03 &     85.27 &     14.86 &     99.09 &      2.20 &      0.22 \\ 
    353K04 &     68.05 &     15.18 &    106.59 &      2.20 &      0.23 \\ 
    353K05 &     77.21 &     14.67 &    103.89 &      2.20 &      0.39 \\ 
    353K06 &     50.70 &     18.49 &    116.00 &      2.20 &      0.38 \\ 
\hline
    545K01 &     34.94 &     18.48 &    136.70 &      2.20 &      0.04 \\ 
\hline
\end{tabular}
  \caption{Bolometer model parameters as described in
    Eq.~\ref{eq:resist} and \ref{eq:conduct} for each
of the Archeops bolometers. \label{tableboloparam}}
\end{table}

Table~\ref{tableboloparam} lists the parameters of all thermometers and
bolometers that were used during KS3 flight (bolometers ordered by channel,
217K05 was blind during the flight).  The differential conductivity at 100 mK
is \( G=dP_{C}/dT=\beta g/T_{100}\) (from Eq.~\ref{eq:conduct}).  These
constants are consistent with those measured on cosmic rays.  A more detailed
description of the previous issues is given in Sect.~\ref{boloresponse}.  The
heat capacity was simply taken as \( C=\tau_{1} G \) where \( \tau_{1} \) is
the first time constant of the bolometer.  Time constants are derived from a
fit on Jupiter data taken during the flight, including electronic filtering
and a Gaussian beam (see Sect.~\ref{sec:timeconstants}).

The differential conductivity \( G\) is taken at \( 100\mathrm{mK} \).
Note that the {}``Kiruna{}'' bolometers have typical conductivity
between 60 and 80 pW/K and heat capacity of 0.3 to 1 pJ/K 
although some of them deviate significantly from this range. 

\subsection{Detector noise}\label{ss:noise}




\begin{figure*}[tb]
\caption{From top to bottom and from left to right are shown the power spectra (in 
  $10^{-17}\,\mathrm{W.Hz^{-\frac{1}{2}}}$) of the Archeops 143K03, 217K06,
  353K01 and 545K01 bolometers, respectively, as a function of frequency (in
  Hz). For comparison we overplot (smooth curve) the detector noise
  contribution for each bolometer as given by the model presented in the
  text.\label{fig:noisemod}}
\end{figure*}

We include in the detector noise model contributions from the FET electronics,
the Johnson noise and the bolometer thermodynamic noise (\cite{Mather:1984}).
To the detector noise we add quadratically the photon noise deduced from our
photometric model.  Note that the bolometer noise is not white at high
frequency due to the bolometer time response.  Figure~\ref{fig:noisemod} shows
an example of power spectra of the time ordered data of four representative
Archeops bolometers during the KS3 flight.  We overplot the noise model
discussed above which is in qualitative agreement without any parameter tuning
at frequencies higher than a few Hz.  We observe an increase of power with
decreasing frequency mainly due to the low frequency systematics.  Although we
have smoothed out the power spectrum we still can observe peaks which
correspond to the sky signal at the spin frequency harmonics which are mainly
dominated by Galactic and atmospheric emissions.  Centered at 1~Hz and in
particular for the high frequency channels there is a very peculiar structure
which may be of atmospheric origin.  Finally at high frequency we observe
correlated structures.  A more detailed description of systematics and their
subtraction is given in Sect.~\ref{sect:syste}.

\begin{table*}
\begin{center}
\begin{tabular}{c|ccc|ccc|cc}
\hline \hline
 Bolometer & $I$ & $R $ & Resp  & $P_{exp}$  & $P_{abs}$  & Eff & NEPphot  & NEPtot   \\
           & $\mathrm{(nA)}$  & $(\Omega)$ & $\mathrm{(10^8\,V/W)}$ & $\mathrm{(pW)}$ & $\mathrm{(pW)}$ & &$\mathrm{(10^{-17}\,W.Hz^{-1/2})}$ & $\mathrm{(10^{-17}\,W.Hz^{-1/2})}$ \\

\hline
    143K01 &      0.57 &      2.78 &      4.98 &       1.7 &    2.6 &       1.3 &      2.2 &      5.1 \\
    143K03 &      0.57 &      2.76 &      4.49 &       1.7 &    2.9 &       1.8 &      2.4 &      3.6 \\
    143K04 &      0.57 &      4.11 &      6.65 &       1.7 &    1.8 &       0.7 &      1.9 &      3.0 \\
    143K05 &      0.57 &      2.33 &      4.27 &       1.7 &    2.9 &       1.7 &      2.4 &      5.1 \\
    143K07 &      0.57 &      2.04 &      3.51 &       1.7 &    3.4 &       1.6 &      2.5 &      6.7 \\
    143T01 &      1.13 &      3.90 &      4.57 &       1.3 &    1.9 &       1.3 &      1.9 &      4.4 \\
    217K01 &      1.70 &      0.95 &      2.04 &       2.0 &    6.1 &       1.3 &      4.2 &      9.1 \\
    217K02 &      1.70 &      0.64 &      1.31 &       2.0 &    8.6 &       2.0 &      5.0 &      9.4 \\
    217K03 &      1.70 &      1.08 &      2.22 &       2.0 &    5.2 &       0.2 &      3.9 &      7.3 \\
    217K04 &      1.22 &      1.11 &      1.93 &       3.5 &    6.7 &       1.8 &      4.4 &      9.5 \\
    217K05 &      1.70 &      0.89 &      1.64 &       2.0 &    7.7 &       1.0 &      4.7 &      7.4 \\
    217K06 &      1.13 &      0.98 &      1.86 &       3.5 &    6.3 &       1.7 &      4.3 &      8.1 \\
    217T04 &      0.91 &      4.39 &      5.29 &       2.1 &    0.5 &       0.7 &      1.2 &      4.4 \\
    217T06 &      0.87 &      5.05 &      4.65 &       2.1 &    2.8 &       1.2 &      2.8 &      6.4 \\
    353K01 &      0.85 &      3.31 &      5.00 &       1.1 &    2.0 &       0.7 &      4.3 &      3.8 \\
    353K02 &      0.85 &      3.98 &      5.37 &       1.1 &    1.9 &       0.6 &      4.2 &      3.8 \\
    353K03 &      0.85 &      3.75 &      5.30 &       1.1 &    1.7 &       0.6 &      4.0 &      3.4 \\
    353K04 &      0.85 &      3.82 &      5.36 &       1.1 &    1.6 &       0.7 &      3.8 &      4.8 \\
    353K05 &      0.85 &      3.42 &      4.99 &       1.1 &    1.9 &       0.7 &      4.2 &      3.7 \\
    353K06 &      0.85 &      5.16 &      5.56 &       1.1 &    3.0 &       0.6 &      5.3 &      4.9 \\
    545K01 &      1.13 &      0.77 &      0.76 &       7.5 &   18.0 &       1.2 &     11.4 &     15.9 \\
\hline
\end{tabular}
\end{center}
\caption{ For all Archeops bolometers from left to right. Photometric quantities
as representative of night flight values: the current, the resistance, and the responsivity.
Expected absorbed power
from a simple photometric model made of subsystem transmission measurements
presented in section~\ref{simplephotomodel}.
Absorbed power as measured (with 1 pW absolute uncertainty) with
the bolometer model described in section \ref{simplebolomodel}.
Efficiency as the ratio
of Jupiter inflight calibration with calibration from the photometric model.
Photon and total noise.
\label{tableallbolophotoquantities}}
\end{table*}

Finally, we present in table \ref{tableallbolophotoquantities} a summary of
the noise properties of all Archeops bolometers. From left to right we include
representative values within the night flight for the main photometric
quantities of those bolometers: current, resistance, and responsivity.  Next
we provide the expected absorbed power from a simple photometric model made of
subsystem transmission measurements presented in
section~\ref{simplephotomodel}.  Then we give the absorbed power as measured
(with 1 pW absolute uncertainty) using the bolometer model described in
section \ref{simplebolomodel}.  The efficiency is given as the ratio of
Jupiter inflight calibration with calibration from the photometric model.
Noise measurements are given at the bolometer level both for photon and total
noise. The photon noise is within a factor 2 of the total noise as measured in
flight conditions.

\section{Preprocessing}
\label{sec:preprocessing}
In this section, we describe the preprocessing of the Archeops data.  This
includes the demodulation of the raw data and the removal of the parasitic
signal introduced by the readout electronic noise.  We also describe the
linearity correction of the bolometers and the flagging of the data affected
by cosmic rays and noise bursts.

\subsection{Prefiltering}
\label{sec:prefiltering}
        \begin{figure}[t]
        \begin{center}
        \caption{{\rm Top : }Kernel of the digital filter used for demodulation (see text
                for details). {\rm Bottom
            : } Fourier power spectrum of the digital filter compared to a square
            filter,
          to the beam pattern and to the bolometer time constant.}
        \end{center}
        \label{preproc}
        \end{figure}

The data are acquired in total power mode via an AC square wave
modulated bias.  All the modulations are driven by the same clock at
76.3 Hz, leading to an acquisition frequency \mbox{$f_{acq}=152.6\,\rm Hz$.}
The AC square wave modulated bias transforms the data into a series of
alternative positive and negative values.  This induces a peak at the
Nyquist frequency, $f_{acq}/2$, in the Fourier power spectrum of the
bolometer data.  This peak fully dominates the signal and needs to be
removed for demodulation.  This is performed by filtering the data and
for this purpose we have constructed a digital filter with the
following constraints:

\begin{itemize}
\item the transition after the cut-off frequency, taken to be 60 Hz,
      must be sharp for a complete removal of the modulation signal,
\item the ringing of the Fourier representation of the filter
  above the cut-off frequency needs to be below the 2\% level, to
  avoid any possible aliasing.
\end{itemize}

These two constraints lead to a digital filter of 23 points whose
kernel is shown on the top panel of Fig.~\ref{preproc}.  The Fourier
response of the filter is shown on the bottom panel and compared to
that of a simple square filter.  We observe that the high frequency
cutting of the digital filter is much sharper than for the square filter
therefore preserving the
signal better.  For comparison, we also plot the Fourier response of
the bolometer time response and of the beam pattern which determine the
spectral band for the signal.  No signal is therefore removed by the
digital filter.

\subsection{Removal of readout digital noise}
\label{readoutnoise}
As discussed above, the bolometer signal is biased with a square
signal.  The data are then amplified by a digital preamplifier and
buffered and compressed by the on-board computer into blocks.  The
blocks are then recorded.  The compression procedure preserves most of
the signal of interest.  Code 32-bit words at the beginning and end of
each block allows us to check those blocks.  These blocks are of
different sizes depending on the nature of data which may correspond to
the signal from the gyroscopes, the bolometers, the thermometers or the
stellar sensor.  The length of the bolometric and thermometric blocks
is of 72 samples.


\begin{figure}[ht] \begin{center}
    \caption{Top:  Fourier power spectrum of KS3 143K01 bolometer data showing
the frequency peaks produced by the readout electronic noise.
     Bottom: Same after preprocessing. The amplitude of the peaks
is significantly reduced.
      \label{fig_pattern}}
\end{center} \end{figure}

While the on-board computer deals with in-flight commands, the data
recording is delayed and a few data blocks are buffered before
recording.  Small offset variations in the electronics lead to
significant differences between the mean value of the last recorded
blocks and the next ones.  As in-flight commands are sent and received
by the on-board computer periodically during the flight (every few data
blocks), the differences in the mean between blocks induce a parasitic
signal on the data.  This parasitic signal shows up in the data as a
periodic pattern of basic frequency $f_{acq}/72$.  Further, as series
of blocks are buffered before recording we also observe in the data
periodic patterns at frequencies which are submultiples of
$f_{acq}/72$.  For most of the bolometers this systematic signal
dominates over the noise and is clearly visible both in the time and
frequency domain.  The top panel of Fig.~\ref{fig_pattern} shows a
zoom-up of the power spectrum of the data of the KS3 143K01 Archeops
bolometer.  We observe on the spectrum a series of peaks which
correspond to the parasitic signal.\\


The subtraction of the parasitic periodic signal can be easily achieved
using a time domain template for it.  Indeed, we have implemented a
fast algorithm for calculating a time varying template of the parasitic
signal.  First of all, for each Archeops timeline we have divided the
data into pieces of N blocks of 720 samples.  The block size
corresponds to the largest period between two in-flight commands.  Then
each piece of data has been reordered into a $720 \times N$ matrix so
that a time evolving pattern of the parasitic signal over 720 samples
can be calculated by smoothing up over the N blocks.  The exact number of
720-samples blocks to be summed up is a compromise between,
first, the minimum signal to noise ratio needed for extracting the
parasitic signal from the data; second, the time evolution rate of the
parasitic signal and third, the minimum time interval needed to
consider that the sky signal varies sufficiently for not contributing
to the template.  We have found that for most Archeops bolometers
$N=100$ is a good compromise.  The constructed template is repeated $N$
times (size in samples of the time interval processed) and then
subtracted from each piece of data.


The bottom panel of Fig.~\ref{fig_pattern} shows the power spectrum of the KS3
143K01 bolometer after applying the above procedure.  The procedure reduces
significantly the peaks.  For example the fundamental frequency peak at
2.12~Hz is reduced to much less than 10 \% of its original value.  The peak at
12.7~Hz, although significantly reduced, is still visible in the preprocessed
spectrum. It will be cut off in the Fourier domain as discussed in
Sect.~\ref{High frequency systematics}.

\subsection{Linearity correction}

The cryostat temperature underwent a slow decrease during the flight,
leading to a slow change of the calibration in $\rm mK/\mu\rm V$.  This
change in calibration can be corrected for by modeling the responsivity of
the bolometer.  Actually, for a TOI $b$ in $\mu\rm V$, we can write the
linearity corrected TOI as follows

\begin{equation}
b_{corr}=-V_{b}\ln\frac{V_{b}+V_{0}-b}{V_{b}}+V_{0}
\end{equation}


The parameters $V_{b}$ and $V_{0}$ are determined from the bias-current curves
of each of the bolometers. After this smooth correction, the calibration
factor in $\rm mK/\mu\rm V$ can be considered as constant over the flight, thus 
allowing for a much easier determination. For the KS3 flight the correction
does not exceed 20~\%, and it is only important for the first 2 hours of flight.
This has been cross-checked via the Galactic plane calibration method
described in Sect.~\ref{sect:galcal}.

\subsection{Flagging of the data}
\label{sect:flag}

In this section, we describe the identification and flagging of
parasitic effects including glitches, noise bursts and jumps in the
data.  For Archeops most glitches are due to the increase of
temperature of the bolometer due to the energy deposited by cosmic rays
hits.  Jumps are essentially due to changes of the equilibrium
voltage of the bolometer and there are only a few during the whole
flight.  Bursts of noise are induced by microphonic noise coming mainly
from the mechanical oscillations of the gondola.  \\

To flag and remove the data affected by the above systematic effects,
the first step is to detect spikes in the TOIs above a certain
threshold level.  For this purpose the r.m.s. noise level, $\sigma^2$,
is estimated locally on a 400 points running window as the standard
deviation from the median value, $m$, of the data after removing 5\% of
the lowest and largest data values.  The data with flux above 8$\sigma$
are considered as glitches.  To preserve the Galactic signal, which
can be sometimes spiky or/and larger than the threshold limit, a
baseline $f_{base}$ fitted as a combination of the two first Fourier
modes is removed whenever data values above the threshold are detected
at Galactic latitude between -10$^{\rm o}$ and +10$^{\rm o}$.  The
value of $\sigma$ is then re-computed and the above criteria
re-applied.  This technique is time-consuming but not required
outside of the Galactic plane where a flat baseline is already a very
good approximation.

The second step is then the flagging of the data.  When the parasitic
signal is due to a glitch, we can model it by the convolution of a
Dirac delta function at time $t_i$ with the sampling window and a
double decreasing exponential function with two time constants
$\tau_{short}$ and $\tau_{long}$.  The first corresponds to the
relaxation time of the bolometer itself and the second is of unknown
origin.  The time constant values depend only on the bolometer and must
be the same for all glitches impacting this bolometer.  The main objective of
this first analysis is not to reproduce faithfully the glitch shape but
to estimate which part of the data is badly affected by it and must be
flagged.  Therefore, the same conservative values are adopted for all
bolometers, $\tau_{short} = 2$~samples (13~ms) and $\tau_{long} =
50$~samples (325~ms).  We then fit the following glitch model

\begin{equation}
  f(t,t_i) = \left[ A_{short}~e^{-\frac{t-t_i}{\tau_{short}}} +
    A_{long}~e^{-\frac{t-t_i}{\tau_{long}}} \right] \ast f_{acq}~~+~~f_{base}
  \label{glitch_model}
\end{equation}

where the free parameters are the amplitudes of exponential functions
$A_{short}$ and $A_{long}$ and the baseline and where $f_{acq}$ is the
sampling frequency.  All data samples within $t_{min}$ and $t_{max}$
given by

\begin{eqnarray*}
  t_{min} & = & t_i - 11\\
  t_{max} & = & t_i
  + \tau_{short} \ln \left[\frac{A_{short}}{0.1\sigma}\right]
  + \tau_{long} \ln \left[\frac{A_{long}}{0.1\sigma}\right]
  + 11
\end{eqnarray*}

are flagged.  So data samples for which the glitch contribution is at a
level higher than 10~\% of the local noise are flagged.  The extra
11~samples margin on each side of the glitch position accounts for the
effect of the digital filter at 23~points discussed in the previous
subsection.  An additional margin of 100~samples is used when the fits
is of poor quality.  This extra flagging concerns about 20~\% of the
glitches detected on the OMT bolometers, 33~\% on the Trapani-like
bolometers and less than 15~\% for the others bolometers of the KS3
flight.\\

Detailed statistics of the number of glitches detected in the KS3
flight are reported in Tab.~\ref{glitch_stat}.  The bolometers from
the Trapani flight are quite sensitives to glitches, 15 to 20 glitches
per min.  Polarized OMT at 353~GHz shows a rate of $\sim$4 glitches per
minute, whereas at 217~GHz and 545~GHz we detect less than 2 glitches
per minute.  At 143~GHz the bolometers present a glitch rate between
1.5 and 4 per minute.  The glitch rate is fully related to the
effective surface of the bolometer which varies between bolometers.  A
larger glitch rate can be explained by a larger effective area of the
spider-web absorber.

\begin{table}[!ht]
  \center
  \begin{tabular}{c|cc}
    \hline \hline
    Bolometer & \# glitches [per min.] & data flagged [\%]\\
    \hline
    143K01 &  1.8 &  0.93 \\
    143K03 &  3.6 &  1.58 \\
    143K04 &  4.2 &  1.74 \\
    143K07 &  1.6 &  0.77 \\
    143K05 &  2.2 &  0.94 \\
    143T01 & 16.8 &  8.58 \\
    \hline       
    217K01 &  1.0 &  0.44 \\
    217K02 &  1.1 &  0.54 \\
    217K03 &  1.3 &  0.55 \\
    217K04 &  1.6 &  0.79 \\
    217K05 &  1.3 &  0.58 \\
    217K06 &  1.5 &  0.79 \\
    217T04 & 16.9 &  8.43 \\
    217T06 & 20.7 & 11.62 \\
    \hline       
    353K01 &  4.8 &  2.15 \\
    353K02 &  4.1 &  1.88 \\
    353K03 &  5.7 &  2.55 \\
    353K04 &  3.8 &  1.72 \\
    353K05 &  4.9 &  2.22 \\
    353K06 &  3.3 &  1.52 \\
    545K01 &  1.1 &  0.76 \\
    \hline
  \end{tabular}
  \caption{Statistics of glitches per minute for the KS3 flight and proportion of
    flagged data.\label{glitch_stat}}
\end{table}

In the above procedure bursts of noise on the data are assimilated to glitches
and the flagging obtained is poor.  To ensure a better flagging, we proceed to
a visual inspection of the data.  We check all the pieces of data found above
the threshold limit and extend manually the flagging if necessary.  Those data
samples affected by noise bursts are flagged as such.  We also observe jumps
on the data which are caused by extreme changes on the DAC currents of the
bolometer.

The values of the DAC currents are stored as housekeeping data and allows us
to correct the data from those jumps via a simple destriping algorithm.  In
addition, by visual inspection we determine the data samples which are
affected by jumps and they are manually flagged.

At the end of the process, a total of 1-2.5~\% (resp.  2-4~\% and
12-18~\%) of the data are flagged for the KS3-like bolometers (resp.
for the OMT and Trapani-like bolometers). Flagged data are then
replaced by constrained realization of noise as discussed in
Subsect.~\ref{sec:noisesimulations}.  \\




\section{Pointing reconstruction \label{sec:pointing}}

The knowledge of the pointing attitude was not needed during the flight but
its accurate {\it a posteriori} reconstruction is critical for mapping
correctly the sky signal. The pointing of each of the detectors in the focal
plane is computed as follows.  First, a pointing solution for the payload is
obtained from the processing of optical data collected by the fast stellar
sensor (FSS) during the flight.  Finally, we estimate the pointing offset with
respect to the payload axis for each bolometer using the reconstruction of the
focal plane from measurements of point sources (see Sect.~\ref{boloresponse}
for details).



We have developed an algorithm to extract star candidates from the
FSS time-sampled photodiode signals (see Sect.~\ref{sec:attitudecontrolsystem}). 
Each star candidate is kept into a
table including its detection time, its position along the diode array and
the electrical intensity observed.  The position of the star candidate
along the diode array is given in terms of an effective diode number.
The electrical intensity measured is proportional to the logarithmic
value of the flux of the star. 

\subsection{Coordinate system definition}

In the following, we use equatorial coordinates \{$\alpha,\delta$\}
to define the position of celestial objects on the sky.  The FSS data
are also easily handled with local coordinates associated with the
gondola frame, for which the zenith corresponds to the gondola spin
axis direction. The direction of a star on the celestial sphere is
then given by $\theta$, the angular distance between the spin axis and
the direction of the star (hereafter the axial distance), and by
$\varphi$, the phase corresponding to the azimuth measured from the
North. 

To reconstruct the pointing direction of the gondola we need to find
the direction of the center of the diode array.  The instantaneous
pointing solution is fully described by the set \{$\alpha_p$,
$\delta_p$, $\varphi$\}, where $\alpha_p$ and $\delta_p$ are the
equatorial coordinates of the gondola spin axis and $\varphi$, the
phase for the diode array.  Note that the phase value $\varphi$ is the
same for all the diodes in the array, therefore also the same for all
detected stars, because the diode array is placed perpendicularly to
the scanning direction. In other words, the number of the diode lightens is only
given by the axial distance of the observed star.  

\subsection{Reconstruction Method}

The goal is to produce an optimal pointing solution as a function of
time.  The reconstruction is based upon the comparison between FSS data
and a dedicated star catalog compiled from the Hipparcos catalog.  The
electrical intensity of stars in the catalog is computed by taking into
account the FSS spectral response.  Hereafter, we call {\it signal} a
star candidate in the list produced by the FSS software and {\it star}
an object taken from the star catalog.   
First of all, we find in the star
catalog the best star to be associated which each FSS signal.  We call
identified signal a signal for which this association is performed.  Via
these associations we obtain a pointing solution for each identified
signal.  Finally, we fit the overall set of identified signals
through the scan path to get a pointing solution as a function of time.
 
\subsection{Initial pointing estimate}

To be able to associate signals to stars, a first estimate of
the pointing solution is needed.  This is obtained via the GPS data which give
the local vertical direction, which corresponds to the spin axis
direction \{$\alpha_p$, $\delta_p$\} to an accuracy of $\sim 1$
degree, taking into account the gondola average pendulation.  Then, we
match signal and star directions and try to identify for each signal
the corresponding star.  There is no direct measurement of the FSS
phase $\varphi$.  We need to reconstruct it from the rotation period
by integration.

\subsubsection{Rotation period}
\label{FSS:rot_per}

We now describe the gondola motion relative to the celestial sphere.
We thus call rotation period the elapsed time between two successive
detections of the same star after one revolution.  Each revolution
takes about 30 seconds.  Due to the Earth motion, the spin axis moves
about 5' in $\alpha$ per revolution.  Each star can thus be seen
several times by the FSS. For each signal, we look for every compatible
signal seen in the preceding revolution.  A compatible signal has a
similar intensity and a nearby diode number.  Time differences between
the signal and those seen in the last revolution are binned into an
histogram.  The most populated bin gives us the rotation period.
Figure~\ref{per} shows the evolution of the rotation period as a
function of time for 3 hours of the KS3 flight.  This evolution is
mainly due to the presence of strong stratospheric winds during the
flight.

\begin{figure}[!hb]
     \begin{center}
       \caption{\label{per} \it{       
           Rotation period evolution during the KS3 flight.
         }}
     \end{center}
\end{figure}

\subsubsection{Star Sensor Phase}

We reconstruct the FSS phase by integrating the angular speed
$\frac{1}{T}$, where $T$ is the rotation period.  The resulting
estimate $\hat{\varphi}$ differs from the phase by a slowly varying
offset.  To correct from this bias, we compare, for each revolution, the
phase of the most intense signals with the phase of the brightest stars
located in the 1.4 degree wide band scanned by the diode array during a
revolution.

\begin{figure}[ht]
        \begin{center}
\caption{\it{
Evolution of the distribution of phase differences 
between signals and bright stars for the KS3 flight.
}\label{Geom1}} 
\end{center}
\end{figure}
\begin{figure}[ht]
\begin{center}
\caption{\it{
Distribution of the axial distance of bright stars versus the diode
number of the corresponding intense signals.  Notice the strange
behavior of the diode 26.  This diode is excluded from analysis.
}\label{Geom2}} 
\end{center}
\end{figure}

The analysis of phase differences $\varphi^{\star}$-$\hat{\varphi}$
gives us the FSS phase offset shown in Fig.~\ref{Geom1}.  The
distribution of axial distance, $\theta$, values of bright stars
associated to intense signals allows us to adjust the geometrical
relation between the axial distance and the diode coordinates along the
array. Figure~\ref{Geom2} shows the distribution of the axial distance
of bright stars as a function of the diode number of the corresponding
intense signal in the FSS. We observe for each diode number the
distribution has a well defined peak from which we can reconstruct the
axial distance for each diode.  The width of the peak is due to the
pendulation motion of the gondola.

\subsection{Star-signal matching algorithm}

The association algorithm used above is based on a comparison of the star and
signal directions.  An error $\vec{\delta}$ on the spin axis direction
($\alpha_{p}$, $\delta_{p}$) translates into a local rotation and then an
error on the reconstructed direction for each of the signals.  The gondola
pendulation is a slow time varying function on scales of a few degrees.
Therefore, the error $\vec{\delta}$ and the local rotation parameters, are
slowly varying functions too. In other words, signals detected within a few
degrees area are thus shifted a roughly equal amount from their true position
on the celestial sphere.

The matching algorithm is based on the above statement and
proceed as follows.
First, for each signal, we associate stars and
signals with compatible positions and intensities.  Second, given a
reference signal, we check whether for the $N$ following signals there
are N stars such that the corresponding shifts are close.  If so, this
displacement is the signature of a local rotation induced by a wrong
reconstruction of the spin axis direction or by a wrong estimate of the
FSS phase.

Free parameters like the number $N$ of signals used or the tolerance on
the angular distance between the signal and its corresponding star
directions can be tuned to optimize the association efficiency.  In
practice, tight cuts on those parameters reduce the probability of
wrong associations but at the same time reduce the number of good
associations available on the regions where the pointing reconstruction
is bad.  To improve this situation we use the fact that the FSS sees a
given star during several revolutions.  Once a good association is
obtained we propagate this information to the whole data set using our
estimate of the rotation period and thus we can improve the association
efficiency and therefore the pointing solution.

\subsection{Pointing solution improvement} \label{PointAlgo}

The axial distance $\theta$ is the only quantity which can be directly
measured.  When signals have been associated to catalog stars, it gives
a way to reconstruct the spin axis direction.  As the position of the
signal and that of its associated star must be the same, the spin axis
is therefore located on a cone centered on the star with an opening
angle $\theta$.  Using two couples, signal-star, we can find the
direction of the spin axis.  Indeed the intersection of the two cones,
one for each couple, leads to two solutions.  Only one of them is
geometrically relevant.  Using the whole data set we can thus correct
the estimate of the spin axis direction during the flight.  We upgrade
the FSS phase taking into account the new estimate of the spin axis
pointing.  The process is iterative to obtain a more accurate estimate
of the pointing for the whole flight.  The increase in accuracy at each
iteration allows us to use tighter cuts to get a better quality
matching between stars and signals.

The FSS dataset available is mainly composed of faint stars making the above
iterative solution very important. Further, calibration uncertainties on the
signal get broader as the intensity decreases.
The associations for the brightest stars allows us to recalibrate the FSS
signals. Adding finer constraints on the intensity of the signal increases the
quality of its association to a star in the catalog. This also improves the
final pointing solution.


\subsection{Scan path fit}

Once the signal-star associations are obtained we have a discrete
pointing solution at the times where the signal were observed. 
Our purpose is to generate an optimal continuous scan path and then we have
to interpolate the pointing solution along FSS data. This solution
should not only be interpolated, but also optimized along the data set $ \{
x_{i},t_{i}\}$.  To get an optimal pointing solution we have to reconstruct
($\tilde{\alpha}_p(t)$, $\tilde{\delta}_p(t)$, $\tilde{\varphi}(t)$),
from the $\{\alpha_{p\:i},\: \delta_{p\:i}, \:\varphi_i \}_i$ set,
where $i$ labels a given signal-star association. This is performed by
computing first a smooth solution for the pointing and then correcting
it.

\subsubsection*{Smooth pointing solution}

We first produce a smooth solution for the pointing ($\tilde{\alpha}_p^{0}$,
$\tilde{\delta}_p^{0}$ and $\tilde{\varphi}^{0}$) by fitting
the set $\{\alpha_{p\:i},\: \delta_{p\:i}, \:\varphi_i \}_i$ 
using a chi-square minimization.  
As the set $\{\alpha_{p\:i},\: \delta_{p\:i}, \:\varphi_i \}_i$  
is irregularly sampled in time we obtain a generic interpolation, 
$\tilde{x}(t)$, of the pointing solution 
through a decomposition of the form

\begin{eqnarray}
\tilde{x}(t) = \sum_{k} c_{k} U(t-\hat{t}_{k})  \label{Udev}
\end{eqnarray}

where each $U(t-\hat{t}_{k})$ is a generic kernel of the
form $sinc(\frac{\pi t}{p})e^{-\frac{t^2}{2 p^2 {\sigma}^2}}$
centered at $t=\hat{t}_{k}$. We choose the parameter $p$
to optimize the representation of the low frequency components
in $\tilde{x}(t)$ and $\sigma$.

The coefficients $\{c_{k} \}$ are obtained from the minimization of the chi-square
$$
\chi^{2}=\sum_{i} 
\left(
x_i - \tilde{x}_i 
\right)^{2}
$$
leading to the following linear system
$$
\sum_{k=1}^N 
\left( 
\sum_{i}  U(t_i-\hat{t}_{l})U(t_i-\hat{t}_{k})
\right)
c_k 
$$
$$
= 
\sum_{i} x_i U(t_i-\hat{t}_{l})
\qquad \textrm{\ with } \:\:\:\: k,l   = 1,2,\ldots,N.
$$

We solve this system for the three quantities of interest $\alpha_{p}(t)$,
$\delta_{p}(t)$, and $\varphi(t)$.

\subsubsection*{Corrected pointing solution}         

Once we have a first smooth solution for the pointing
($\tilde{\alpha}_p^{0}$,$\tilde{\delta}_p^{0}$ and $\tilde{\varphi}^{0}$) we
compute corrections to it $\Delta\tilde{\alpha}_p(t)$,
$\Delta\tilde{\delta}_p(t)$ and $\Delta\tilde{\varphi}(t)$.  For this purpose
we decompose these 3 quantities in terms of kernel functions as in
Eq.~\ref{Udev}.  We call $\Delta a_{k}$, $\Delta d_{k}$ and $\Delta p_{k}$ the
decomposition coefficients for $\Delta\tilde{\alpha}_p(t)$,
$\Delta\tilde{\delta}_p(t)$ and $\Delta\tilde{\varphi}(t)$ respectively.  In
this case we consider high frequency terms to optimize the pointing solution.

The FSS dataset $\{\alpha_{p\:i},\: \delta_{p\:i}, \:\varphi_i \}_i$ can be
rewritten more explicitly as $ \{\alpha_{p\:i},\: \delta_{p\:i}, \:\varphi_i,
\:\theta_i, \:\alpha_i^{\star}, \:\delta_i^{\star} \}_i $.  $\theta_i$ is a
linear function of the diode number.  $\alpha_i^{\star}$ and
$\:\delta_i^{\star}$ are the coordinates of the star corresponding to signal
$i$.  This set can also be expressed for the star position in gondola frame
coordinates $\{\alpha_{p\:i},\:\delta_{p\:i}, \:\varphi_i, \:\theta_i,
\:\varphi_i^{\star}, \:\theta_i^{\star} \}_i $.  We can obtain an estimate of
the pointing corrections by comparing the reconstructed star positions with
the pointing position at the time of their observation.

\begin{equation}
\chi^{2}=\sum_{i} \left\lbrace 
\left(
\frac{{\varphi_i^{\star}} - {\tilde{\varphi}_i}}{\sigma^{\varphi}_i}
\right)^{2} + \left(
\frac{{\theta_i^{\star}} - \theta_i}{\sigma^{\theta}_i}
\right)^{2}
\right\rbrace
\label{chi2corrections}
\end{equation}

We note $\sigma^{\theta}_i$ and $\sigma^{\varphi}_i$ the errors
associated with the measurements of $\theta$ and $\varphi$ obtained in
the previous section.  There are two sources of asymmetry between
$\sigma^{\theta}_i$ and $\sigma^{\varphi}_i$.  The axial distance
coordinate $\theta$ is directly measured by the FSS. The phase
coordinate must be reconstructed, once the spin axis direction is
known.  The second source is geometric.  A diode covers 1.9 arcminutes
in the cross-scan direction by 7.6 arcminutes along the scan.  We use
$\sigma^{\theta}_i$ = $\sigma$ and $\sigma^{\varphi}_i$ = $2\sigma$.

The star coordinates $\varphi^{\star}$ and $\theta^{\star}$ in the
gondola frame depend on the spin axis direction.  A variation
$\Delta\tilde{\alpha}_p$ and $\Delta\tilde{\delta}_p$ in this
direction induces a modification of the coordinates $\theta^{\star}$
and $\varphi^{\star}$.  To first order, we have
$$
\left\{
         \begin{array}{l@{\quad}l}
{\theta^{\star}}={{\theta}^{\star}}^o+c_{11}\Delta\tilde{\alpha}_p + c_{12}\Delta\tilde{\delta}_p & \\ 
{\varphi^{\star}}={{\varphi}^{\star}}^o+c_{21}\Delta\tilde{\alpha}_p + c_{22}\Delta\tilde{\delta}_p.  &  \\
        \end{array}
     \right.
     $$
The coefficients $c_{ij}$ are known functions of ${\alpha_{p}}^{o}$,
     ${\delta_{p}}^{o}$, ${\theta^{\star}}^o$ and
     ${\varphi^{\star}}^o$.  

Then Eq.\,\ref{chi2corrections} becomes

\begin{eqnarray*}
\chi^{2}=\sum_{i} 
\left(
\frac{
{
\varphi_i^{\star \: o}} - {\tilde{\varphi}^o_i}
+c_{21}\Delta\tilde{\alpha}_p + c_{22}\Delta\tilde{\delta}_p
-\Delta\tilde{\varphi}
}
{\sigma^{\varphi}_i}
\right)^{2} \\
+ \left(\frac{{\theta_i^{\star \: o}} - \theta_i
+c_{11}\Delta\tilde{\alpha}_p + c_{12}\Delta\tilde{\delta}_p }
{\sigma^{\theta}_i}
\right)^{2}.
\end{eqnarray*}

To achieve convergence, we follow the same steps as the method
described in \ref{PointAlgo}.  We first calculate the correction on
the spin axis direction $\Delta\tilde{\alpha}_p(t)$ and
$\Delta\tilde{\delta}_p(t)$.  This leads us to minimize the quantity

\begin{eqnarray}
\chi_1^{2}=\sum_{i} 
\left(
\frac{
{
\varphi_i^{\star \: o}} - {\tilde{\varphi}^o_i}
+c_{21}\Delta\tilde{\alpha}_p + c_{22}\Delta\tilde{\delta}_p
}
{\sigma^{\varphi}_i}
\right)^{2} \\
+ \left(\frac{
{
\theta_i^{\star \: o}} - \theta_i
+c_{11}\Delta\tilde{\alpha}_p + c_{12}\Delta\tilde{\delta}_p
}
{\sigma^{\theta}_i}
\right)^{2}.
\label{chi1theta}
\end{eqnarray}
The phase correction is then obtained by taking

\begin{equation}
\chi_2^{2}=\sum_{i} 
\left(
\frac{
{
\varphi_i^{\star \: o}} - {\tilde{\varphi}^o_i}
+c_{21}\Delta\tilde{\alpha}_p + c_{22}\Delta\tilde{\delta}_p-\Delta\tilde{\varphi}
}
{\sigma^{\varphi}_i}
\right)^{2}.
\qquad 
\label{chi2phase}
\end{equation}

The minimization of $\chi_2^{1}$ and $\chi_2^{2}$ leads to the iterative
resolution of two linear systems with free parameters $\Delta a_{k}$,
$\Delta d_{k}$ and $\Delta p_{k}$.  \\

In the above we have assumed that the photodiodes array was oriented
perpendicularly to the pointing direction.  This hypothesis can be
verified by comparing the phase for the stars with the phase of the FSS
as a function of the diode number.  This comparison is shown on 
Fig.~\ref{residus}.  We observe a phase shift which indicates that the
photodiodes array is tilted along the scan direction.  Given the
1.8\,m
focal length of the parabolic mirror and a 1mm photodiode area along
cross-can direction, we find an inclination of $\sim 3$ degrees.  The
phase of each signal is thus corrected to take this effect into
account.

\begin{figure}
\begin{center}
\caption{ 2D histogram showin`g a variation of phases difference
between signals and associated stars with the signal diode number.
\label{residus}} 
\end{center}
\end{figure}

\subsection{Accuracy}

We have of two independent but complementary ways of assessing the
accuracy of the Archeops pointing reconstruction.  A first estimate can
be obtained from the distribution of coordinate differences between the
signals and their associated stars.  Figure~\ref{err:contgoni}
show the distribution of errors in the plane
axial distance-phase before and after scan path fit respectively.  The
95\% and 68\% confidence level contours are displayed in white.  We
observe that the axial distance coordinate has intrinsically a better
accuracy by a factor 2.5.  Further, we notice a significant decrease of
the errors for both the axial distance and the phase.

We can also estimate the errors in the pointing reconstruction via the
Fisher matrix of the free parameters in the scan-path fit described by
Eq.\ref{chi1theta} and \ref{chi2phase}.  This gives us a
continuous estimate of the pointing error which is used to flag those
regions for which the pointing is badly known.  Hereafter, we call this
flag on the data bad pointing flag.
The distribution of equatorial coordinate differences in
Fig.~\ref{err:cont} shows that the attitude reconstruction is achieved
with an accuracy better than 1.5 and 1 arcmin in
RA and DEC respectively, at the 1-$\sigma$ level.

\begin{figure}[h!]
   \begin{center}
\caption{
\label{err:contgoni}
From top to bottom, distribution of errors in axial distance - phase plane
with 95\% and 68\% confidence levels (in white) before and after scan path fit
respectively.}
\end{center}
\end{figure}


\begin{figure}[ht!]
  \begin{center}
    \caption{
      \label{err:cont}
      KS3 flight 95\% end 68\% confidence levels for error distribution in
      equatorial coordinates after scan path fit.}
\end{center}
\end{figure}

\section{Bolometer response} 
\label{boloresponse}
\begin{figure}[!ht]
  \begin{center}
    \caption{Top: map of Jupiter for the 143K03 bolometer in $\mu$V before time
      constant deconvolution. The map is represented in the par and cross scan
      direction in arcmin. Bottom: As above but after deconvolution from the
      time constant.
\label{mapjupiterbeforeafterdeconv}}
  \end{center}
\end{figure}

We describe in this section the reconstruction of the Archeops focal
plane parameters for the KS3 flight. For this purpose we estimate 
the time response of the bolometers, the optical response of
the photometric pixels and 
the focal plane geometry on the celestial sphere. The
focal plane is reconstructed using planets observations. The
brightest one, Jupiter, is observed twice at $\sim$~16.5 and
21h00~UT hours and can be considered as a point source at the Archeops
resolution (apparent diameter of 45 by 42~arcsec).  We also use Saturn
observations obtained at 15h36 and 18h427~UT hours to cross check the results.
Saturn can be also considered as a point source at the Archeops
resolution (apparent diameter of 19 by 17~arcsec).  \\

For each detector, we start by computing local maps of the
planets in azimuth-elevation coordinates which correspond to the
along-scan and cross-scan directions.  These maps are obtained by
projecting the TOI data without filtering.  To remove the zero level
in these maps we estimate a baseline in the TOI which is then
subtracted.  The latter is estimated from a TOI where all the flagged
data are interpolated using a constrained realization of noise.
The TOI signal obtained for planet observations is the superposition of
two main effects.  First, the convolution of the source sky signal with
the beam pattern of the photometric pixels.  Second, the convolution of
the bolometric signal with the time response of the bolometers which is
characterized by a time constant.  Both effect are clearly visible in
the Jupiter map shown on the top panel of
Fig.\,~\ref{mapjupiterbeforeafterdeconv}.  The beam pattern convolution
widens up the point source signal both in the cross-scan and along-scan
directions.  The effect of the time response convolution appears as a
tail in the map along the scan direction.

In our analysis we first estimate the bolometer time constants for
each using in-scan profiles of the Jupiter or Saturn map.
Then we deconvolve the TOI from the bolometer time constant and
recompute local maps as the one presented on the bottom panel
of Fig.\,~\ref{mapjupiterbeforeafterdeconv}. From these maps, 
we characterize the beam pattern of the photometric pixels.

\subsection{Time response}
\label{sec:timeconstants}
\subsubsection{Optical time constants estimate}

The time response, TR, of the bolometers can be described by the
combination of two decreasing exponentials with time constants
$\tau_{1}$ and $\tau_{2}$

\begin{equation}
  TR(t) = (1-\alpha)~e^{-t/\tau_{1}} + \alpha~e^{-t/\tau_{2}}
\end{equation}

with $\alpha$ a mixing coefficient to be estimated from the data. As Archeops
scans the sky at roughly constant speed, the effect of beam pattern and time
response are degenerate in the along-scan direction.  In order to have the
simplest possible model, but that allows us to separate both effects, we will
assume that the beam pattern shape is symmetric along the scanning direction.
\footnote{Notice that for a constant angular speed the effect of the
  time response convolution can be assimilated to a beam convolution.} \\

The time constants are estimated fitting the Jupiter profiles using a
$\chi^2$ minimization for a grid of 3 parameters $\tau_1$, $\tau_2$ and
$\alpha$ which are set in the range [1,10]~ms, [10,100]~ms and [0,1]
respectively.  The profiles used are the 4~arcmin cross-scan average of
local maps of the two Jupiter crossings.  For each set of parameters,
we deconvolve the initial TOI from the $TR(\tau_{1},\tau_2,\alpha)$.
We then fit a Gaussian on the rising part of the beam profile and use a
Gaussian with the same amplitude and sigma for the decreasing part.  We
do not account for the higher part of the beam shape that can present
several maxima (especially multimode ones) by fitting only the lower
80\% of the profile data.

We compute the minimum of the $\chi^2$ in the ($\tau_{1}$, $\tau_{2}$),
($\tau_{1}$, $\alpha$) and ($\tau_{2}$, $\alpha$) planes.  The best-fit
parameter values are obtained from the average of the two maxima
obtained for each parameter.  By integrating the surface we obtain
directly the 1$\sigma$ error bars.  If $\alpha$ is compatible with 0
within 1$\sigma$, we re-compute the estimation of $\tau_1$ using a
single time constant model to reduce the error bars.

Figure~\ref{beamprofile} shows one of the Jupiter map profile for the
217K04 bolometer before (in red) and after (in black) deconvolution
from the bolometer time response.  The tail in the profile due to the
time response is clearly suppressed after deconvolution.

\begin{figure}[ht]
  \begin{center}
    \caption{217K04 Beam profile on Jupiter before (in red) and after (in black) deconvolution of the
      two time constants ($\tau_1 = 5.57_{-1.08}^{+1.01}$~ms, $\alpha =
      0.48_{-0.04}^{+0.04}$ and $\tau_2 = 38.38_{-4.20}^{+3.80}$~ms).}
    \label{beamprofile}
  \end{center}
\end{figure}

\begin{table}[hhh]
  \begin{center}
    \begin{tabular}{c|c|c|c}
      \hline \hline
      bolometer  &   $\tau_1 (ms)$ & $\alpha$ & $\tau_2 (ms)$ \\
      \hline
      143K01  &   $  6.87^{+ 0.78}_{-0.83}$  & $ 0.32^{+ 0.05}_{-0.06}$  & $ 62.16^{+ 35.03}_{-24.23}$ \\
      143K03  &   $  5.98^{+ 0.52}_{-0.58}$  & $ 0.00$  & - \\
      143K04  &   $  7.36^{+ 0.97}_{-0.99}$  & $ 0.20^{+ 0.09}_{-0.10}$  & $ 21.84^{+ 19.71}_{-11.84}$ \\
      143K05  &   $  6.91^{+ 0.82}_{-0.79}$  & $ 0.25^{+ 0.08}_{-0.08}$  & $ 21.49^{+  6.39}_{ -5.54}$ \\
      143K07  &   $  6.23^{+ 0.76}_{-0.85}$  & $ 0.00$  & - \\
      143T01  &   $  4.94^{+ 0.39}_{-0.40}$  & $ 0.00$  & - \\ \hline
      217K01  &   $  6.07^{+ 1.58}_{-1.93}$  & $ 0.38^{+ 0.07}_{-0.06}$  & $ 23.20^{+  9.34}_{ -6.64}$ \\
      217K02  &   $  5.57^{+ 1.10}_{-1.34}$  & $ 0.00$  & - \\
      217K03  &   $  0.52^{+ 2.20}_{-0.52}$  & $ 0.00$  & - \\
      217K04  &   $  5.57^{+ 1.01}_{-1.08}$  & $ 0.48^{+ 0.04}_{-0.04}$  & $ 38.38^{+  3.80}_{ -4.20}$ \\
      217K05  &   $  5.81^{+ 0.93}_{-0.92}$  & $ 0.00$  & - \\
      217K06  &   $  7.04^{+ 0.53}_{-0.55}$  & $ 0.00$  & - \\
      217T04  &   $  6.57^{+ 0.61}_{-0.54}$  & $ 0.00$  & - \\
      217T06  &   $  0.00^{+ 0.00}_{-0.00}$  & $ 0.00$  & - \\ \hline
      353K01  &   $  2.28^{+ 1.12}_{-1.28}$  & $ 0.00$  & - \\
      353K02  &   $  1.17^{+ 1.69}_{-0.17}$  & $ 0.00$  & - \\
      353K03  &   $  2.26^{+ 1.00}_{-1.11}$  & $ 0.00$  & - \\
      353K04  &   $  2.14^{+ 0.87}_{-1.05}$  & $ 0.00$  & - \\
      353K05  &   $  3.79^{+ 0.99}_{-1.38}$  & $ 0.00$  & - \\
      353K06  &   $  3.25^{+ 0.96}_{-1.19}$  & $ 0.00$  & - \\ \hline
      545K01  &   $  0.28^{+ 0.54}_{-0.28}$  & $ 0.00$  & - \\
      \hline
    \end{tabular}
    \caption{Bolometer time constants for the Archeops KS3 flight\label{results_cte}}
  \end{center}
\end{table}

In Tab.~\ref{results_cte} we present, for each of the Archeops bolometers,
the values of $\tau_1 (ms)$,  $\alpha$ and $\tau_2 (ms)$ 
obtained from the analysis of the Jupiter profiles. The analysis
of the Saturn profiles provides consistent results.

\subsubsection{Estimate of the bolometer time constant from glitches}

The time response of the bolometer can also be estimated using the
signal from cosmic ray glitches with short time constant (see
Sect.~\ref{sect:flag}) which hit directly the bolometer.  To a very
good approximation, for these glitches, the signal is just the
convolution of a Dirac delta function with the bolometer time response
and the sampling kernel and therefore they have all the same shape.  A
template of this can be obtained by piling up all short glitches in
the data after common renormalization both in position and amplitude.
We can then fit to this template the glitch model
(Eq.~\ref{glitch_model}) to estimate $\tau_{short}$.  Notice that only
a few glitches have an additional significant long time constant
preventing us from reconstructing an accurate template.

Figure~\ref{glitch_template} shows in black the glitch template
computed for the bolometer 217K01.  In red we trace the best glitch
model fit for it, corresponding to a time constant of
$7.8\pm0.11$~ms.
 
\begin{figure}[!ht]
  \begin{center}
    \caption{Glitch template for the bolometer 217K01. A single time constant
      model have been fitted to the data. For the best fit, traced in red, the time
      constant is $7.8\pm0.11$~ms.\label{glitch_template}}
  \end{center}
\end{figure}

In Fig.~\ref{time_glitch} the values of $\tau_{short}$ obtained from
the fit of glitches are compared to the shorter optical time constant
$\tau_1$ estimated using Jupiter (Tab~\ref{results_cte}). These are
compatible within 1$\sigma$ for a large number of detectors.  The
observed discrepancies may be due to the intrinsic degeneracy between
time response and beam pattern and the way we break it.  Equally, we
can imagine different time delays in the detector response depending on
where exactly the glitch hits.

\begin{figure*}[!tb]
  \begin{center}
    \caption{Comparison between glitch and Jupiter short time constant estimates.\label{time_glitch}}
  \end{center}
\end{figure*}

The second time constants $\tau_{long}$ and $\tau_2$ differ by at least one
order of magnitude.  The long time constant measured on glitches can be
interpreted as a longer thermal link for glitches which hit the immediate
surrounding of the bolometer and will not be considered in the following.  By
contrast, the second optical time constant, $\tau_2$, must be taken into
account to accurately deconvolve from the bolometer time response.

\subsection{Optical response}

\subsubsection{Beam pattern model}
After deconvolving the Archeops TOI from the bolometer time response, we
reconstruct local maps of Jupiter to estimate the beam pattern shape. The beam
patterns for the Archeops photometric pixels happen to be asymmetric in
particular for the multimode systems (all the 217 GHz detectors but 217K01,
217K02 and 217K05 and the two 545GHz detectors).

We model the main beam shape for each photometric pixel using the {\it
  Asymfast} method, described in \cite{asymfast}. This method is based on the
decomposition of the main beam shape into a linear combination of circular 2D
Gaussian. This allows us to accurately and simply represent asymmetric beams
and to convolve full sky maps with them in a reasonable amount of time. This
is of great interest when producing simulations to estimate the beam transfer
function in the spherical harmonic plane (\cite{tristram_cl}).

The Archeops main beams have been modeled using up to 15 Gaussians. The
residuals after subtraction of the model from the Jupiter local maps are less
than 5\%.  Figure~\ref{asymfast} shows an example of this multi-Gaussian beam
modeling for the photometric pixel 143K03 using 7 circular 2D Gaussians.  From
top to bottom and from left to right we represent the beam pattern shape from
the Jupiter local maps, its multi-Gaussian fit, the residuals and the
histogram of the residuals (black line). The latter are shown to be compatible
with Gaussian distributed noise (red line in the figure).

\begin{figure}[!ht]
  \begin{center}
    \caption{From top to bottom and from left to right, for the
    photometric pixel 143K03, the beam pattern map in $\mu$V from Jupiter
    observations, its multi-gaussian decomposition using seven 2D
    Gaussians, the residuals and their distribution which is compatible
    with Gaussian distributed noise (red line). \label{asymfast}}
  \end{center}
\end{figure}

The resolution for each of the photometric pixels has been estimated from a 2D
elliptical Gaussian fit to the local beam maps from Jupiter observations. The
FWHM values given in Table~\ref{fwhm_arch} are the geometric mean value in the
two directions. We present as well the ellipticity as computed from the ratio
between the minor and the major axis. For monomode systems the beamwidth is
about 11~arcmin. The multimode systems at 217~GHz and 545~GHz have larger
beams with FWHM $\sim$15~arcmin. The 353~GHz beams are monomode but are by
construction illuminating a small part of the primary mirror to have clean
polarized beams. A degradation of the beams has been noticed in the KS3
flight compared to ground-based measurements (which are close to the diffraction limit)
due to a slight out-of-focus position of the secondary mirror after the crash landing of the KS2 flight.  

\begin{table}[!ht]
  \center
  \begin{tabular}{cccccc}
    \hline \hline
    Bolometer & fwhm [arcmin]& e  & Modes \\
    \hline
    143K01 & 11.0 & 0.76 & S \\
    143K03 & 11.7 & 0.74 & S \\
    143K04 & 10.9 & 0.85 & S \\
    143K05 & 11.9 & 0.78 & S \\
    143K07 & 11.7 & 0.80 & S \\
    143T01 & 10.2 & 0.87 & S \\
    \hline
    217K01 & 12.0 & 0.72 & S \\
    217K02 & 11.8 & 0.73 & S \\
    217K03 & 17.5 & 0.99 & M \\
    217K04 & 15.9 & 0.75 & M \\
    217K05 & 11.3 & 0.79 & S \\
    217K06 & 15.1 & 0.72 & M \\
    217T04 & 14.0 & 0.65 & M \\
    217T06 & 15.2 & 0.80 & M \\
    \hline
    353K01 & 11.9 & 0.78 & S \\
    353K02 & 12.0 & 0.79 & S \\
    353K03 & 11.9 & 0.78 & S \\
    353K04 & 12.0 & 0.77 & S \\
    353K05 & 12.1 & 0.71 & S \\
    353K06 & 12.2 & 0.72 & S \\
    \hline
    545K01 & 18.3 & 0.68 & M \\
    \hline
  \end{tabular}
  \caption{Resolution in terms of the FWHM and ellipticity of the beam pattern
    for the Archeops photometric pixels in the KS3
    flight. See text for details. The optical mode of
    the bolometers is indicated in the last column: S for single
    mode bolometers and M for multimode bolometers. \label{fwhm_arch}}
\end{table}

No specific treatment have been
applied in the data analysis to account for far side lobes
as from Planck optical modelization they are expected to be
at the percent  level.

\subsubsection{Focal plane reconstruction}

The position of each photometric pixel in the focal plane with respect
to the Focal Plane Center (FPC) is computed using Jupiter
observations.  This then allows us to build the pointing of each pixel
using the pointing reconstruction as described in Sect.~\ref{sec:pointing}.
In practice we use the relative positions of the 2D circular Gaussians
of the {\it Asymfast} decomposition to estimate the center of the beam
in focal plane coordinates.
Fig.~\ref{plan_focal} shows the reconstruction of the Archeops focal plane in
azimuth (along scan) and elevation (across scan) coordinates.  The Archeops
focal plane is about 2~degrees high and 2.5~degrees wide.
\begin{figure}[!ht]
  \begin{center}
    \caption{Focal plane of Archeops reconstructed using Jupiter observations.\label{plan_focal}}
  \end{center}
\end{figure}





\section{Description and subtraction of systematics}
\label{sect:syste}

In this section we describe in details the main systematic effects on the
Archeops data as well as the methods and algorithms used for their
subtraction.  Because of the circular Archeops scanning strategy the sky
signal shows up on the data at frequencies which are multiples of the spin
frequency.  This naturally leads to three distinct regimes in frequency.
First, the very low frequency components at frequencies well below the spin
frequency ($f< 0.01$~Hz) which are mainly due to $1/f$-like noise and
systematics.  Second, the spin frequency components ($0.03 < f < 3$~Hz) at
frequencies which contain most of the sky signal of interest.  And finally the
high frequency components at frequencies much larger than the spin frequency
($f > 10$~Hz) which are dominated by detector noise.


\begin{figure*}[!tb]
  \begin{center}
    \caption{Left column: from top to bottom, raw Archeops data in
    $\mu$V (black curve) and reconstructed very low frequency drift
    (red curve) for the 143K03, 217K06, 353K01 and 545K01 bolometers
    respectively.  Right column: from top to bottom, Archeops data in
    $\mu$V after removal of the very low frequency parasitic signals
    for the 143K03, 217K06, 353K01 and 545K01 bolometers respectively.
    \label{fig:vlf}}
  \end{center}
\end{figure*}

\subsection{Very low frequency systematics}
\label{sec:verylowsyste}

At very low frequency the Archeops data are dominated by systematics
coming mainly from temperature fluctuations of the three cryogenic
stages at 100~mK, 1.6~K and 10~K and from the variation of air mass
during the flight due to changes in the balloon altitude.  The left
column of Fig.~\ref{fig:vlf}, from top to bottom, shows the
raw Archeops data for the 143K03, 217K06, 353K01 and 545K01 bolometers
in the period from 14h00 to 29h00 UT time.  We observe a very low
frequency drift in the data which is very well correlated within
bolometers and also with the low frequency components in the
housekeeping data from thermometers placed at each of the cryogenic
stages and with measurements of the altitude of the balloon.

This drift is removed via a decorrelation analysis using as templates
the housekeeping data described above and a fifth order polynomial
defined in the time interval of interest.  To compute the correlation
coefficients we first smear out and undersample both the Archeops data
and the templates, and then, we perform a linear regression.  A final
estimate of the drift is obtained from the best-fit linear combination
of the templates which are previously smoothed down to keep only the
very low frequency signal.  In the left column of Fig.~\ref{fig:vlf}
we overplot in red the reconstructed very low frequency drift for the
four Archeops bolometer. In the right column of the figure we show the
Archeops data after subtraction of the low drift estimate which reduces
the signal amplitude by a factor of 50.  Although the decorrelation
procedure is very efficient, we can still observe a correlated low
frequency parasitic signal between bolometers.  To avoid the mixing up
of the bolometer signals at this stage of the processing of the data,
this effect is considered within the map making pipeline described in
Sect.~\ref{The Archeops maps}.

\subsection{Spin frequency systematics}
\label{Spin frequency}

\begin{figure*}[!tb]
  \begin{center}
    \caption{Left column: from top to bottom, Archeops data in $\mu$V
      (black curve) and reconstructed spin frequency systematics (red curve,
      including the CMB dipole and atmospheric emission) for the 143K03,
      217K06, 353K01 and 545K01 bolometers respectively.  Right column: from
      top to bottom, Archeops data in $\mu$V before (black curve) and after
      (red curve) removal of the spin frequency systematics for the 143K03,
      217K06, 353K01 and 545K01 bolometers respectively.
      \label{fig:lf}}
  \end{center}
\end{figure*}

The Archeops data at the few first multiples of the spin frequency are
particularly important. They contain the large angular scale signal of
the sky whose mapping is one of the main purposes of the Archeops
experiment.  At these frequencies in addition to the CMB anisotropies
and the Galaxy, two main components can be identified, the CMB
dipole and the atmospheric emission.  Although the former is critical
for the calibration of the Archeops data at 143 and 217 GHz (see
Sect.~\ref{dipole}) for this analysis we consider both of them as
systematics.

As above, we perform a decorrelation analysis to remove these parasitic
signals.  A template for the dipole in mK$_{CMB}$ is simulated using
the COBE best dipole solution (\cite{paperdipole}) and the local
velocity vector during the flight.  The atmospheric contamination is
mainly due to the variation of the air mass induced by changes in the
altitude and the pointing elevation of the payload.  As templates for
this effect we use the housekeeping data corresponding to measurements
of the altitude of the balloon and the reconstructed elevation of the
focal plane.  To compute the correlation coefficients, we smear out and
undersample both the Archeops data and the templates and perform a
linear regression.  The final estimate of the spin frequency
systematics are obtained from the best-fit linear combination of
templates.  The left column of Fig.~\ref{fig:lf} shows from top to
bottom, the Archeops data in $\mu$V (black curve) and the reconstructed
spin frequency systematics (red curve) for the 143K03, 217K06, 353K01
and 545K05 bolometers respectively.  We observe that at the low
frequency channels the dipole contribution (exactly at the spin
frequency) dominates while at the high frequency the atmospheric
emission is much more important and dipole becomes negligible.

\begin{figure*}[!tb]
  \begin{center}
    \caption{From left to right and from top to bottom power spectrum 
      in $\mu$V$\sqrt{Hz}$ of 
             the time ordered Archeops data before (black curve) and 
             after (red curve) subtraction of the spin frequency systematics
             for the 143K03, 217K06, 353K01 and 545K01 bolometers.
      \label{spin_freq}}
  \end{center}
\end{figure*}
  
The spin frequency systematic estimate is then filtered out in the
range of frequency of interest, low passed to remove very low frequency
drifts in the templates and high pass to reduce the noise in the
templates, and subtracted from the Archeops data.  The right column of
Fig.~\ref{fig:lf} shows the Archeops data for the four bolometers
before and after subtraction of the spin frequency systematic
estimates.  After subtraction we clearly observe at all frequency the
Galactic signal.  This shows up as peaks at each revolution and of
increasing amplitude with increasing frequency channel.  This is better
shown on Fig.~\ref{spin_freq} where we represent the power spectrum
in $\mu$V$\sqrt{Hz}$
of the Archeops time ordered data before (black
curve) and after (red curve) subtraction of the spin frequency
systematics.  At 143 and 217~GHz the dipole contribution appears clearly
as a peak at the spin frequency before subtraction.  After subtraction
we clearly distinguish the Galactic signal which dominate the spectrum
at 353 and 545~GHz.

Beside the atmospheric signal which is correlated to templates there is
a residual parasitic atmospheric signal.  The latter can be
qualitatively reproduced by simulating turbulent atmospheric layers
drifting across the Archeops scanning beam.  Typical gradients of half
a mK$_{RJ}$ over 10 degrees azimuth are observed at 545~GHz with an
evolving period of 1000 seconds.

%

\subsection{High frequency systematics}
\label{High frequency systematics}

        
\begin{figure*}[!ht]
  \begin{center}
    \caption{From left to right and from top to bottom power spectrum
     in $\mu$V$\sqrt(Hz)$ of the low frequency processed Archeops data before (in black)
     and after (in red) decorrelation from the high frequency noise
     for the 143K03, 217K06, 353K01 and 545K01 bolometers respectively.
      \label{fig:hfdecorrelation}}
  \end{center}
\end{figure*}

The Archeops data present, at frequencies $f>5$~Hz, parasitic noise
which shows up on the time domain as noise bursts which are neither
stationary nor Gaussian.  In the Fourier plane this parasitic signal is
observed in the form of well defined correlated structures or peaks.
This parasitic noise is very likely related to microphonic noise.  Its
contribution to the data was significantly reduced in the KS3 flight
with respect to the KS1 flight by increasing the distance between the
payload and the spinning motor.  The latter was placed far away in the
flight chain with no rigid link between them.
Fig.~\ref{fig:hfdecorrelation} shows in black the power spectra of
the Archeops low frequency processed data for the 143K03, 217K06,
353K01 and 545K01 bolometers.  Structures around 15, 22, 40 and 50 Hz
are clearly visible in the power spectrum.  We can also observe a peak
at 28.5~Hz in the power spectrum of the 353K03 bolometer data.

These structures are also observed on the power spectrum of the data from the
focal plane thermometers and from the blind bolometer. Moreover, they are in
phase correlated, in the temporal domain, with the bolometer ones. This fact
allows us to efficiently remove this parasitic noise via a decorrelation
analysis in the Fourier plane using as templates the housekeeping data for the
focal plane thermometers and the blind bolometer.

Assuming a linear model we can write the Fourier transform of the
parasitic signal for the bolometer $i$ as $B^{par}_{i}(\nu) =
F_{i}(\nu) \times T(\nu)$ where $T$ represent the templates described
above and $F_{i}(\nu)$ is a frequency dependent correlation
coefficient.  As the parasitic signal is both well localized in time
and in frequency, we perform a regression analysis in the Fourier plane
using the windowed Fourier transform over time intervals of size $L$ in
order to compute the correlation coefficient $F_{i}(\nu)$.  To maximize
the efficiency of the algorithm and to limit the subtraction of sky
signal which may be accidentally correlated with the templates we apply
the decorrelation analysis only in a few predefined frequency intervals
where the parasitic signal dominates.  Further to obtain a robust
estimate of the correlation coefficient we average it over $N$ time
intervals so that for each of the frequency intervals it reads

%
\begin{equation}
F_{i}(\nu)=\frac{\sum_{k=0}^{N-1}\tilde{B}_{i}^{k}(\nu)\tilde{t}^{k^{*}}(\nu)}
{\sum_{k=0}^{N-1}\tilde{t}^{k}(\nu)\tilde{t}^{k^{*}}(\nu)}.
\end{equation}

where $\tilde{f}$ represents the Fourier transform of function $f$.

The estimate of the high frequency parasitic noise is then subtracted
from the data in the Fourier plane. We repeat this analysis for each of
the available templates. 
The choice of the parameters of the method, i.e $L$, $N$, results from
a trade-off : we want to minimize the parasitic power in the TOI
spectrum, but we also want that the CMB signal remained unaffected by
the process.  Simulations leads to a best choice of parameters
$L=32768$ samples and $N=32$ to get a CMB power reduced by less than
1\%.


In Fig.~\ref{fig:hfdecorrelation} we plot the power spectra for the
bolometers 143K03, 217K06, 353K01 and 545K01 before and after
subtraction of the high frequency parasitic signals.  We observe that
the parasitic signal is significantly reduced in particular at the
lowest frequencies.  However for most of the bolometers around $40$~Hz
the processing although efficient in removing the systematic effect is
not satisfactory (residuals are much larger than for lower frequencies)
and therefore in the following the Archeops TOIs are high pass filtered
at $38$~Hz before any scientific analysis.  We also observe that, in
some of the bolometers very well localized peaks in the spectrum are
not removed in our process.  These peaks are manually characterized 
and the
signal at their frequencies is removed in the Fourier domain for all
Archeops bolometers.

\section{Data quality checks and noise properties}
\label{sec:dataquality}

\begin{figure*}[!ht]
  \begin{center}
    \caption{Left column: from top to bottom, time-frequency representation of the 217T04
             bolometer data and of the expected Galactic signal for it. Right column: same for
             the 217K04 bolometer.
      \label{timefreqanalysis}}
  \end{center}
\end{figure*}
As discussed in the previous section the Archeops data are affected by
parasitic microphonic noise at high frequencies and by other
unidentified systematics in the whole frequency range.  None of them
are neither stationary nor Gaussian and their contributions vary
significantly among bolometers.  In this section we briefly describe
how we performed the selection of the best bolometers, in terms of
level of residual systematics and of noise properties, which are used
for the construction of the Archeops sky maps presented in
Sect.~\ref{The Archeops maps}.

\subsection{Time-frequency analysis of the Archeops data}

We have performed a time-frequency analysis of the Archeops data using the
Discrete Wavelet Packet Transform (DWPT) as described in \cite{wavpaper}. The
top left panel of figure~\ref{timefreqanalysis} shows the power distribution
in the time-frequency plane for the data of the 217T04 bolometer.  At low
frequency and because of the particular scanning strategy of Archeops, we can
observe spikes and some broader structures which correspond mainly to Galactic
signal. This is clearly shown by the bottom plot where we trace the expected
Galactic signal for this bolometer. Also at low frequency we distinguish the
atmospheric residuals and the $1/f$-like noise on the bolometer.  We can
further observe from 20 to 40 Hz structures which vary on time being
particularly strong around 21h00 UT time. These structures are residuals,
after subtraction, of the high frequency systematics described before. Notice
that the level of systematics is significantly larger than the high frequency
noise making this bolometer useless for scientific purposes.

On the top right panel of figure~\ref{timefreqanalysis} we present the
time-frequency analysis of the 217K04 bolometer. As above, we can clearly
distinguish at low frequencies the Galaxy contribution whose template is
represented in the bottom right panel. We observe for this bolometer no
intermediate frequency structures as for the previous bolometer.  The
systematic contribution has significantly reduced by the high frequency
subtraction and the residuals are well below the high frequency noise.  This
simple but efficient analysis allows us to exclude from the further processing
those bolometers which present either strong or highly time variable
systematic residuals. Typically the Trapani bolometers presented a large
amount of contamination and were excluded.  We also realized that the noisier
bolometers were in general more affected by residual systematics.

\begin{figure}[t]
\caption{{\it Top}: Average wavelet power spectrum in $\mu$V
of the TOD of the 217K04 Archeops bolometer
as a function of frequency computed from its DWPT {\it Bottom:} Time evolution of the
power spectrum of the TOD of the 217K04 Archeops bolometer computed from its DWPT.}
\label{fig:func_sigmat}
\end{figure}

\subsection{Long term non-stationarity of the noise}
\label{sec:longtermnonstat}

On the top panel of Fig.~\ref{timefreqanalysis}, at very high frequency we can
distinguish a slow decreasing of the noise power with time which is the main
cause of non stationarity in the Archeops data.  This is due, as discussed in
Sect.~\ref{sec:verylowsyste}, to the sudden increase of temperature of the
Archeops cryostat when taking off and to its slow cooling down to the nominal
value of about 95~mK during the flight.  To account for this non stationarity
we have modeled the Archeops data as a time modulated-stationary wavelet
process (\cite{wavpaper}).  These processes correspond to a continuous
generalization of piece-wise stationary processes.  They are described by two
main variables : the mean power spectrum of the data and a time varying
function, $\sigma (t)$, which account for the time variations of the power
spectrum.

Figure~\ref{fig:func_sigmat} shows on the top and bottom panels the
time averaged wavelet spectrum and the $\sigma (t)$ function
respectively for the noise of bolometer 217K04. On the top plot we can
clearly distinguish the increase of noise power with frequency which is
due to the deconvolution from bolometer time response.  On the $\sigma
(t)$ function, there are two distinct regimes corresponding first to a
fast cooling of the cryostat in the first two hours and then to a
roughly constant temperature of the focal plane during the last 10
hours of flight.  Notice that our estimate of $\sigma (t)$ is noisy and
therefore for further processing we generally smooth it.  We can also
observe variations on $\sigma (t)$ above the noise limit for the second
regime, that, although small with respect to those on the first two
hours of flight, need to be taken into account in any further
processing.
 
\subsection{Gaussianity of the noise}
\label{Stationarity and gaussianity}
        

Up to now, we have only considered the power spectrum evolution to define
the level of stationarity of the data.  To be complete in our analysis
we have first to characterize the Gaussianity of the noise distribution
function and second to check its evolution with time.  In practice and
to reduce the uncertainties in the analysis it is more convenient to
proceed the other way around.

We have implemented a Kolmogorov-Smirnov test in the Fourier domain to
check the evolution with time of the noise distribution function of
each of the bolometers.  Notice that the intrinsic bolometer noise can
be considered Gaussian to a very good approximation and therefore any
changes on the distribution function of the noise will indicate the
presence of significant residuals from systematics.  We work in the Fourier
domain both to speed up the calculations and to isolate the noise,
which dominates at intermediate and high frequencies, from other
contributions like the Galactic and atmospheric signals.  We have
performed the test in consecutive time intervals of about 7 minutes
which we compare two by two.  We divided the frequency space in bins of
1~Hz.  The test is considered to fail when the probability of having a
greater Kolmogorov-Smirnov statistics under the equal distribution
hypothesis is less than 1\%.

%
The results of the test for the KS3 bolometers 143K03, 217K06, 353K06 and
545K01 are shown on Fig.~\ref{kstest}.  The white points represent failing
intervals in the time-frequency domain.  Between 16h00 and 27h00 UT, the test
is successful except for the lowest frequency bins (f$\leq$10~Hz), in which
the Galactic and atmospheric signals, which are neither Gaussian nor
stationary, dominate.  We notice that elsewhere the number of points where the
test fails is no more than 1\% of the total, that is, exactly what is expected
under the hypothesis of no time evolution of the distribution function.
Moreover, they do not exhibit any clustering.  Around 28h00 UT, the sunrise on
the gondola induces strong time evolution due to the heating of the 10~K stage.  \\

\begin{figure}[!ht]
  \begin{center}
    \caption{From top to bottom, results of the Kolmogorov-Smirnov test on the bolometers
         143K03, 217K06, 353K06 and 545K01 respectively. The white polygons correspond to intervals in
      the time frequency plane where the test is considered to fail (see text for details).
      \label{kstest}}
  \end{center}
\end{figure}

In the range from 16h00 to 27h00 UT, as the properties of the distribution
function does not vary, we can globally study the Gaussianity of the noise.
For this purpose we have implemented a simple test in the
Fourier domain by fitting a Gaussian to the histogram of the coefficients of the Fourier
decomposition of the time ordered data which were binned in 1~Hz intervals as above
and computing the $\chi^{2}$ value for the fit.  
We consider the test to fail if for Gaussian distributed data the probability of having a greater
reduced $\chi^{2}$ that the one measured is significantly below 5 \%. 
Figure~\ref{ajustgaussfig_perbinks3} shows the results of the test for the KS3
bolometers, 143K03, 217K06, 353K06 and 545K01.  We trace the reduced
$\chi^{2}$ as a function of the frequency bin. In dark and light blue we
overplot the $\chi^{2}$ values for which the probability of having a larger
one considering a Gaussian distribution are 95 \% and 5 \% respectively.  The
reduced $chi2$ measured are almost everywhere below or around the 5 \% limit,
except in the first frequency bin where the Galactic signal, highly non
Gaussian, dominates. We can therefore consider that the data are in a first
approximation compatible with Gaussianity.

\begin{figure*}[!tb]
  \begin{center}
    \caption{From left to right and up to bottom, maximum reduced $\chi^2$ (97 d.o.f) of 
      the Gaussian fit of the histogram of the Fourier coefficients of the
      Archeops data for the 143K03, 217K06, 353K06 and 545K01 bolometers
      respectively. In dark and light blue we overplot the $\chi^{2}$ values
      for which the probability of having a larger one considering a Gaussian
      distribution are 95 \% and 5 \% respectively.  (see text for details on
      the analysis)
     \label{ajustgaussfig_perbinks3}}
  \end{center}
\end{figure*}

\subsection{Noise power spectrum estimation and simulations}
\label{sec:noisesimulations}

We have performed noise simulations, which we call constrained realizations of
noise, to fill the gaps in the data.  In this case we use a simple algorithm.
First, we reconstruct the low frequency noise contribution via interpolation
within the gap using an irregularly sampled Fourier series.  Finally, we
compute the noise power spectrum locally (in time intervals of about 1 hour
around the gap) at high frequency and then we produce a random realization of
this spectrum.  Notice that we are only interested in keeping the global
spectral properties of the data.  Moreover, the gaps are in general very small
in time compared to the piece of the data used for estimating the power
spectrum and therefore this simple approach is accurate enough.\\

For the first estimate of the CMB angular power spectrum with the Archeops
data (\cite{archpaper}) we needed an accurate estimate of the noise angular
power spectrum.  For this purpose, we have estimated the time power spectrum
of the noise for each of the Archeops bolometers using the algorithm described
in \cite{amblardhamilton}.  This algorithm relies on the iterative
reconstruction of the noise by subtracting in the TOD an estimate of the sky
signal.  The latter is obtained from a coadded map which at each iteration is
improved by taking into account the noise contribution.  From the
reconstructed noise timeline we can then obtain for each bolometer both the
average noise power spectrum and the $\sigma(t)$ function described in
Sect.~\ref{sec:longtermnonstat}.  Since the noise in the Archeops data can be
considered as Gaussian distributed (Sect.~\ref{Stationarity and gaussianity})
these two quantities are enough to simulate noise timelines using the
algorithm presented in~\cite{wavpaper}.  The fake timelines can then be
projected into maps for the estimate of the noise angular power spectrum.  In
general a few hundreds simulations are needed to obtain a reasonable estimate.

\subsection{The best bolometers}

From the above analysis of the Archeops noise we have chosen for each
frequency band those bolometers for which the residual systematics are
well below the noise level.  For the CMB channels at 143~GHz and
217~GHz the selected bolometers (143K03, 143K04, 143K05, 143K07, 217K04
and 217K06) correspond to the more sensitive ones going from 94 to 200
$\mu K \ s^{1/2}$.  At 353~GHz the bolometers were all of similar
quality and we keep all of them.  At 545~GHz only one bolometer was
available, 545K01.

\section{Calibration}
\label{sec:calibration}
We describe in this section the global absolute calibration and intercalibration
of the Archeops data.  The former is performed using three different
types of calibrator : the CMB dipole, the Galaxy and the planets
Jupiter and Saturn.  At low frequencies (143 and 217~GHz) the dipole is
the best absolute calibrator.  At higher frequencies we need to
use the Galaxy because the dipole signal is too faint with respect to
the noise and systematics.

\subsection{CMB dipole}\label{dipole}

At low frequencies, the CMB dipole is a very good absolute calibrator
(\cite{piat:2003}, \cite{Cappellini:2003}) and therefore it
constitutes the primary absolute calibration of the Archeops data.
Here we use the total dipole which is the sum of the solar dipole
(constant in time) and the Earth induced dipole (with annual variations
due to the Earth change of velocity) computed at the time of flight.
Indeed, the dipole calibration has the following advantages:
1) the dipole is spread over the all sky and thus it is always present
whatever the pointing, 2) it is much brighter (typically a factor of
100) than the CMB anisotropies, but still faint enough so that non
linearity corrections are usually not needed, and finally 3) it has the
same electromagnetic spectrum as the CMB anisotropies so that no color
corrections need to be applied.  The only drawback is that we must
assume a constant response for the instrument throughout a wide range
of angular scales, {\sl i.e.} an extrapolation from $\ell=1$ to 1000.
The dipole being an extended source we need to account for the beam and
the spectral transmission of each of the detector to generate the point
source calibration.  \\

In the Archeops case, the dipole signal is expected to contribute only to the 
fundamental rotation frequency $f_{\mathrm{spin}}$.
%
%
However, the drift of the rotation axis during the flight leads to a
broadening of the $f_{\mathrm spin}$ dipole contribution which needs to
be taken into account.  Furthermore, as discussed in Sect.~\ref{Spin
frequency}, other signals show up at the spin frequency including both
Galactic and atmospheric signals.  To account for these, we
compute the total dipole calibration coefficients from the correlation
analysis described in Sect.~\ref{Spin frequency}.  We actually
produced templates of the Galactic emission using 
extrapolation of the dust emission to the Archeops frequency 
(\cite{finkbeiner}) and templates for the atmospheric emission 
using the housekeeping data from
the altitude and elevation of the balloon.  The template for the total
dipole in time domain was obtained by deprojecting, using the Archeops
pointing, a simulated CMB dipole map for which the dipole
amplitude and direction were taken from the WMAP results (\cite{wmap}).
The left panel of Fig.~\ref{dipole_maps} shows the simulated total
dipole map in Galactic coordinates for the Archeops coverage and
centered in the Galactic anticenter.  On the right panel we plot the
Archeops reconstructed CMB dipole for the 143K03 bolometer.  We
observe a very good agreement between the two maps but for some stripes
when crossing the Galactic plane and on the North at high Galactic
latitude.


\begin{figure*}[t]
  \begin{center}
    \caption{From left to right, simulated and reconstructed CMB dipole maps for 
the Archeops 143K03 bolometer centered at the Galactic anticenter. 
To reconstruct both dipole maps the timelines
have been band pass filtered. This introduces discontinuities on the maps
due to the scanning strategy. 
 \label{dipole_maps}}
  \end{center}
\end{figure*}

Indeed, the CMB dipole is detected with a signal to noise better than 500 for
12 hours of data on the 143 and 217~GHz channels.  The errors on the dipole
calibration coefficients come mostly from systematic effects.  We produced
several versions of the calibration coefficients by changing the templates
used in the fit.  We noticed that the result is very stable with respect to
even the most extreme cases.  From those tests, we can deduce that the overall
uncertainty ($1\ \sigma$) is 4~\% (resp.  8~\%) for the 143 (resp.  217)~GHz
photometric pixels.  The larger uncertainty for the 217~GHz channel reflects
the spectrum of the atmospheric contamination. Notice that the dipole
calibration is performed at the early stages of the analysis before removing
the spin frequency systematics (see section~\ref{sect:syste}).  When applied
to the 353~GHz channel, there is a residual contaminant at the same level as
the dipole.  Considering the 545~GHz channel in the analysis can help in this
case and allows finding the dipole calibration coefficient at 353~GHz in
agreement with the Galactic calibration coefficients within 20~\% as described
below.

\subsection{The Galaxy}
\label{sect:galcal}

At high frequencies 353 and 545~GHz, the Galactic emission has to be used to
calibrate the Archeops data.  The best data, in term of spectral coverage and
absolute calibration accuracy, are the FIRAS spectra (\cite{Mather:1990}).
FIRAS, on board the COBE satellite, is a scanning Michelson interferometer
that has provided low-resolution spectra in a low (2 - 20 cm$^{-1}$) and high
(20 - 100 cm$^{-1}$) frequency band, for 98.7\% of the sky.  For individual
pixels, the signal-to-noise ratio of the FIRAS spectra is about 1 at high
Galactic latitude and $\sim$50 in the Galactic plane. The FIRAS maps used here
were obtained by fitting each FIRAS spectrum using a modified Black Body with
a $\nu^{\beta}$ emissivity law and extrapolating the fit at Archeops
frequencies.  Since we are searching for the best representation of the data
and not for physical dust parameters, we restricted the fit to the frequency
range of interest (this avoids the need for a second dust component, of the
type proposed by \cite{finkbeiner}).  The FIRAS brightness maps are then
converted to photometric maps with the flux convention of constant $\nu I_{\nu}$. To be
compared to the FIRAS maps and to maintain the best possible photometric
integrity, the Archeops data are convolved by the FIRAS beam and put into the
FIRAS Quadrilateralized Spherical Cube (QSC).  At this stage, we have Archeops
maps at the FIRAS resolution that can be directly compared to the FIRAS
maps at the Archeops frequencies to derive the Galactic calibration factors.  \\

Due to its high signal-to-noise ratio and its extension, the Galactic plane is
the best place to derive the calibration factors.  We compute Galactic
latitude profiles of both maps ($|b|<30^o$) at selected longitudes and perform
a best straight-line fit to the Archeops-FIRAS profile correlation from which
we derive the calibration factor and its error bar.  An example of this fit is
shown on Fig.  \ref{fig_calib_gal} where we observe a linear correlation
between the Archeops and FIRAS profile intensities.  The whole calibration
process gives calibration factors with statistical errors of about 6 $\%$.
Details on the whole calibration process
are given can be found in \cite{lagadou_calib}.\\

\begin{figure}[ht] 
\begin{center}
\caption{FIRAS/Archeops Galactic profiles correlation on the Galactic plane 
(bolometer 353K01 at 353~GHz). Fitting a straight-line gives the calibration factor.
\label{fig_calib_gal}}
\end{center} 
\end{figure}

Our procedure has been extensively tested.  We have applied the calibration
scheme to the comparison between the FIRAS and DIRBE data at 140 and 240 $\rm
\mu$m and found results in very good agreement with those of \cite{fixsen}.
We also tested our procedure on the \cite{finkbeiner} maps, although these
maps exclude the Galactic plane below $|b|$=7$\rm ^o$.  We obtain gain
differences which are less than 6\% across the Archeops frequencies.  We also
used Archeops maps obtained using different low-frequency filters to test the
effect of the filtering on the calibration. We found that
it modifies the calibration factors by only $\sim$3$\%$.\\

Although the calibration on the Galaxy is less accurate for the low-frequency
than for the high-frequency channels of Archeops, we compare the CMB dipole
and Galaxy calibration factors on Fig.~\ref{fig_compa_dipole_gal}. We observe
a good agreement between both calibrations, with a mean ratio of the
calibration factors lower than 1.05 and all of them compatible with 1 within
the 1$\sigma$ error bars. This clearly demonstrates the robustness of the
methods used for the extended emission calibration and the consistency of the
data reduction from TOIs (dipole calibration) to maps (Galactic calibration).

\begin{figure}[ht]
  \begin{center}
    \caption{
      Comparison of the Galactic and Dipole calibration factors 
(in $\mathrm{mK}/\mu\mathrm{V}$).
      The error bars (both statistical and systematic) are of
      about 4\% and 8\% for calibration on the dipole at 143 and
      217~GHz respectively and from 6\% to 15\% for the calibration on the Galaxy.
      \label{fig_compa_dipole_gal}}
  \end{center}
\end{figure}

\subsection{Jupiter calibration}
We have also performed a point source calibration using the measurements
at the two independent crossings of the planets Jupiter and Saturn (5
times fainter than Jupiter). 
For this purpose we proceed as follows: 1) We obtain the beam pattern
shape from the Jupiter measurements after deconvolution from the
bolometer time response 2) We compute the flux of the sources (in
$\mu$V) using pixel photometry up to a radius of 40~arcmin starting
from the center of the beam pattern.  3) Finally, we compare the fluxes
to a model of the temperature emission of the sources
(\cite{Moreno:1998}) that reproduces radio observations of Jupiter and
Saturn from 20 to 900~GHz.  Taking into account uncertainties on the
model, absolute errors on point source calibration with Jupiter are
estimated to be $\sim$12\%.

The ratio between Jupiter and Saturn flux measurements at a given frequency
does not depend on the absolute instrument calibration.  We find ratios of
$0.97\pm0.016$ and $1.02\pm0.020$ at 143 and 217~GHz respectively which are
not compatible with the one measured by \cite{goldin:1997} ($0.833\pm0.012$)
at similar frequency (171~GHz).  Due to the large brightness of Jupiter (about
1~K$_{RJ}$ equivalent brightness) we expect some non-linearity on the
bolometer response which could be the cause of this difference.  This problem
in addition to the uncertainties on the knowledge of the beam pattern and in
particular of the far side lobes could also explain the small discrepancies
between the Jupiter calibration and the dipole calibration as shown on
Fig.~\ref{calib_jup}.

\begin{figure}[ht]
  \begin{center}
    \caption{
      Comparison of the point source (Jupiter and Saturn) and dipole
      calibration factors. The error bars are about 4\%
      and 8\% for calibration on the dipole at 143 and 217~GHz
      respectively and about 12\% for the calibration on the sources
      (essentially due to the uncertainty of the thermal emission
      model).
      \label{calib_jup}}
  \end{center}
\end{figure}

\subsection{Intercalibration}
\label{sec:intercalib}

\begin{table}[!ht]
  \center
  \begin{tabular}{ccc}
    \hline \hline
    Bolometer  & relative calibration ($\mu$V/$\mu$V) & error (stat.)\\
    \hline
    143K01 & 1.16 & 2.1\%  \\
    143K03 & 1    &   -  \\
    143K04 & 1.63 & 2.3\%  \\
    143K05 & 1.15 & 1.8\%  \\
    143K07 & 1.39 & 1.9\%  \\
    143T01 & 2.06 & 2.3\%  \\
    \hline
    217K01 & 1.80 & 1.1\% \\
    217K02 & 1.72 & 0.8\% \\
    217K03 & 9.09 & 2.5\% \\
    217K04 & 0.903 & 0.7\% \\
    217K05 & 3.21 & 1.2\% \\
    217K06 & 1    &  - \\
    217T04 & 1.50 & 1.1\% \\
    217T06 & 1.02 & 1.02\% \\
    \hline
    353K01 & 1    &    - \\
    353K02 & 1.13 & 0.68\%  \\
    353K03 & 1.19 & 0.70\% \\
    353K04 & 1.02 & 0.70\% \\
    353K05 & 1.20 & 0.79\% \\
    353K06 & 1.05 & 0.78\% \\
    \hline
  \end{tabular}
  \caption{Relative calibration coefficients and their relative
  statistical errors from Galactic profiles at constant Galactic
  longitude.  \label{tab:intercal}}
\end{table}

The polarization signal for experiments like Archeops is reconstructed,
in a first approximation, from the differences between pairs of
bolometers (see Sect.~\ref{sec:polarmaps}).  Therefore the accuracy
on this reconstruction is very sensitive to the relative calibration
between bolometers.
In the case of the Archeops 353~GHz polarized channels, the absolute
calibration on the Galaxy presented above is not accurate enough for the
direct reconstruction of the polarized maps. Thus, we have implemented a
relative calibration algorithm based on the inter-comparison, for all
bolometers in the same channel, of Galactic profiles at constant Galactic
longitude.  This algorithm has also been applied to the unpolarized channels
at 143, 217~GHz as a cross check of the absolute calibration analysis.

Assuming $N$ bolometers per channel we can measure, $N$ Galactic profiles as a
function of the Galactic latitude, $b$,
$$
s_j(b) = \alpha_j\overline{s}(b)+n_j(b)
$$
where $\alpha_j$ and $n_j(b)$ are the intercalibration coefficient and
noise contribution for the bolometer $j$ respectively, and
$\overline{s}(b)$ is the true Galactic profile at the frequency of
interest.  Our algorithm, which is based on a $\chi^2$ minimization
constrained via Lagrange's multipliers, estimate simultaneously the
intercalibration coefficients and the mean Galactic profile.  This
algorithm is described in details on appendix \ref{app:intercalib}.  We
have performed robustness test to validate the algorithm.  In
particular we have checked that the results are not affected by the
choice of constraint and that the analytical error bars obtained for
the intercalibration coefficients are reliable.  \\

\begin{figure}[ht]
  \begin{center}

    \caption{From left to right, Archeops Galactic profiles in $\mu$V 
      at constant Galactic longitude respectively before and after the
      intercalibration for the 353~GHz bolometers.
    \label{intercalibdataprofiles}}
  \end{center}
\end{figure}

The algorithm was applied to the Archeops data which were previously
processed as described in the previous sections.  The profiles were
obtained by averaging the signal samples within latitude bins.  The
errors are computed assuming that the noise is white, uncorrelated
between bolometers and stationary.  The noise of each sample is
estimated from either data outside the Galactic plane ($|b| \ge
25^\circ$) or high-pass filtered data above 10~Hz (where Galactic
signal is negligible).  Results are very similar (differences between
the two methods are less than 3\%, inducing negligible difference in
intercalibration coefficients) in the two cases.  Finally, care has
been taken on keeping the same sky coverage for all bolometers when
computing the Galactic profiles.

For the polarized channel the presence of strongly polarized regions on
the sky may affect the computation of the intercalibration coefficients.
To avoid this, we proceeded in two steps.  First, we compute the
relative calibration coefficients and mean Galactic profile on the full
common sky area for the six polarized bolometers.  Using these
coefficients we build polarization maps and label strong polarized
areas which are then excluded from the analysis.  We then build again the
Galactic profiles and recompute the relative calibration coefficients.
After two iterations, we observed that the estimates of the
intercalibration coefficients are stable.  We show on
Fig.~\ref{intercalibdataprofiles} the results of this analysis for
the 353~GHz channel.  From left to right we plot the Galactic profiles
before and after intercalibration respectively.  We observe that the
intercalibration is achieved with a high degree of precision.

Final results on the intercalibration coefficients for the Archeops
data are given in Tab.~\ref{tab:intercal} for all the bolometers at
143~GHz, 217~GHz (unpolarized) and 353~GHz (polarized).  The
$1\sigma$ statistical errors on the relative calibration coefficients
are at most 2.5\% and always below 1\% for the polarized channel.
Error bars are smaller at high frequency because the Galaxy signal is
stronger.

\subsection{Overall sensitivity}
\begin{table*}[!ht]
  \begin{center}
    \begin{tabular}{ccccccc}
      \hline\hline
      Freq.(GHz) & $N_{bol}$ &  
      $\mu\mathrm{K_{RJ}/Hz^{1/2}}$  & MJy/sr/Hz$^{1/2}$  &
      $\mu\mathrm{K_{CMB}/Hz^{1/2}}$ & $10^6 y /\mathrm{Hz^{1/2}}$ & $\mu\mathrm{K_{CMB}.s^{1/2}}$ \\
      \hline
      143 &  6  &    50  &  0.031 &    87  &    23  &    61  \\
      217 &  7  &    39  &  0.056 &   127  &   286  &    90 \\
      353 &  6  &    82  &  0.315 &  1156  &   132  &   817 \\
      545 &  1  &    77  &  0.702 &  9028  &   495  &  6384 \\
    \end{tabular}
    \caption{
      The Archeops KS3 in--flight timeline sensitivities per channel.
      The best bolometers on each frequency channel are optimally
      combined ($N_{bol}$) to obtain a sensitivity at the instrument
      level. From left to right the units correspond to a Rayleigh--Jeans spectrum, then a
      constant $\nu I_\nu$ spectrum, and a CMB spectrum. Sensitivity
      to the SZ effect is measured with the dimensionless $y$ Compton
      parameter. To convert to 1~$\sigma$ and to one second integration,
      we can simply divide by $\sqrt{2}$: see an example in the last
      column.
      \label{tab:time_sensitiv}}
  \end{center}
\end{table*}

Using the observed response and white noise level of the bolometers, we
can now estimate the global efficiency of the instrument during the KS3
flight.  Table~\ref{tab:time_sensitiv} gives the instrument
sensitivities in various units at the timeline level.  This places
Archeops within the best range of instrumental instantaneous
sensitivities for millimetre continuum measurements.  We also compute
the averaged sensitivity, quoted in Tab.~\ref{tab:map_sensitiv}, by
using a total integration time of 12 hours and a total sky coverage of
30~\%.  In this case Archeops is not within the best experiments
because the sensitivity per pixel is diluted by the large covering area
needed for reconstructing the large angular scales on the sky.

\begin{table*}[!ht]
  \begin{center}
    \begin{tabular}{ccccccc}\hline\hline
      Freq.(GHz) & $N_{bol}$ &  
      $\mu\mathrm{K_{RJ}}$  & MJy/sr/Hz  &
      $\mu\mathrm{K_{CMB}}$ & $10^6 y$ & Jy\\
      \hline
      143 & 6 &    57  &  0.036 &    98  &    27  &    1.2 \\
      217 & 7 &    44  &  0.064 &   144  &   325  &    2.2 \\
      353 & 6 &    94  &  0.358 &  1312  &   150  &   12.1 \\
      545 & 1 &    87  &  0.797 & 10251  &   562  &   27.0 \\
    \end{tabular}
  \end{center}
  \caption{The Archeops KS3 in--flight map sensitivities per
    channel. The best bolometers on each frequency channel are
    optimally combined ($N_{bol}$) to obtain a sensitivity at the map
    level. A square pixel of 20 arcminutes is taken to compute the
    average $1\ \sigma$ noise. KS3 flight roughly covered $30~\%$ of the
    total sky, which represent 110,000 pixels. A bolometer has observed a
    pixel on the map during an average time of 0.4 seconds.
    \label{tab:map_sensitiv}}
\end{table*}

\section{The Archeops sky maps}
\label{The Archeops maps}

In this section we finally describe how we obtain submillimetre sky maps
from the Archeops timelines and pointing information.  
First and prior to projection we remove low frequency
drifts on the Archeops timelines via a destriping algorithm.  Then,
these timelines are processed in three different ways to produce after
projection: CMB, Galactic intensity and polarization maps.

\subsection{Destriping}
\begin{figure*}[!ht]
\begin{center}
{    \begin{tabular}{cc}
\end{tabular}}
\end{center}
\caption{
 Left: Power spectrum of the time ordered data of the bolometer 545K01 before (black)
and after (red) destriping. Right: Zoom-in of the left plot at first multiples of
the spinning frequency. \label{fig-arch-dest-2}}
\end{figure*} 

Even after the subtraction of the very high and low frequency
identified systematic effects from the data, we can observe residual
stripes on the Archeops simple coadded maps.  These are mainly due to
low frequency drifts in the data coming from atmospheric residuals and
other thermal backgrounds.  Notice that there is no electronic $1/f$
component because of the AC bolometer modulation scheme (see
Sect.~\ref{readoutnoise}).  To remove those drifts we have
implemented a destriping algorithm (\cite{bourrachotthesis}) making the
assumption that the scanning direction is generally not related to the
orientation of the structures on the sky. 
To destripe we compute a low frequency baseline in the timelines by
minimizing the ratio
between the rms in the cross-scan and in the in-scan directions
directly from the time ordered data (no reconstruction of maps is
needed).  To represent the baseline we use a basis of localized
functions $U_{k}(t)$ where

\begin{equation}
\begin{array}{rcl}
U(t)     & = & sinc \left(\frac{\pi t}{\Delta} \right) \exp 
         \left( - \ \frac{t^{2}}{2 \Delta^{2} \sigma^{2}}\right) \\
U_{k}(t) & = & U(t-t_{k}) \\
t_{k} & = & k \Delta .
\end{array}
\end{equation}

These functions are regularly sampled and contain only frequencies lower than
$1/(2\Delta)$. 

The minimization is performed outside the Galactic plane over boxes
with sizes related to $\Delta$ (typically a few tens of square degrees).
The cut on the Galactic plane is obtained from a Galactic mask derived
from Galactic template maps computed at the Archeops frequencies
(using \cite{finkbeiner}).  The
algorithm is applied in steps of decreasing $\Delta$ (4000, 1000, 500
and 300) to focus on different frequency ranges.  In any case, data
at frequencies above 1~Hz are not affected by this method.  Prior to
this process we generally apply a classical destriping algorithm based on the
minimization of the variance per pixel in the maps to produce a first
approximation of the baseline and in particular of the lower frequency
components (below 0.7~Hz).

We have applied the destriping algorithm to simulated Archeops data at
217 and 545~GHz including correlated noise at low frequencies and
Galactic and CMB emissions.  No bias has been observed in the
estimation of the Galactic signal.  For the CMB the power is reduced by
at most 5 \%.  The full destriping transfer function in the multipole
space is presented in \cite{tristram_cl}.

An example of the application of the destriping procedure to the
Archeops data is shown on Fig.~\ref{fig-arch-dest-2}.  On the left
panel, we represent the power spectrum of the 545K01 bolometer data
before (black) and after (red) application of the destriping method.
We observe that the noise is reduced significantly at low frequencies
and in particular the spectrum flattens.  Further we can observe from
the right plot that power spectrum signal at the few first multiples of
the spinning frequency are much broader before destriping.  We can
conclude, by comparing the 353 and 545~GHz data, that this extra
structures come mainly from the atmospheric emission.

\begin{figure*}[t!]
  \begin{center}
    \caption{From left to right and from top to bottom MDMC decomposition of the Archeops
data at intermediate frequencies for the 143K03, 217K04, 353K01 and 545K01 bolometers.
The black, blue and red line correspond to power spectrum in arbitrary units of the raw data, 
the parasitic-like and the Galactic-like contributions respectively. \label{smica_ozone}}
  \end{center}
\end{figure*}

\subsection{Specific processing for Galactic maps}
\label{sec:specGalactic}
\begin{figure}[b!]
  \begin{center}
    \caption{Power spectrum in arbitrary units of the parasitic-like component 
      for the MDMC analysis of the bolometer 353K01 for different time
      intervals. \label{smica_ozone_evol}}
  \end{center}
\end{figure}

\subsubsection{Atmospheric contamination}

For producing Galactic maps from the Archeops data we need first to
remove the residual parasitic atmospheric noise.  The destriping
algorithm described above, although very efficient at frequencies lower
than 1~Hz, can not fully eliminate it.  From Fig.~\ref{fig-arch-dest-2}
we observe that the latter shows up on the power spectrum of the time
ordered data as residuals at the spin frequency multiples.  This also
produces two well defined structures, between 0.9 and 1.6~Hz, in the
power spectrum.  This is shown on Fig.~\ref{smica_ozone} where we plot
the power spectrum of the TOI for the 143K03, 217K04, 353K01 and 545K01
bolometers.  Notice that the parasitic noise shows a common spectrum
shape for all the Archeops bolometers and that its total intensity
increases with the frequency of observation.

To estimate this parasitic atmospheric noise and remove it from the
data we have used a modified version of the MCMD-SMICA component
separation algorithm (\cite{smica}) which can work directly on time
ordered data.  We have assumed a very simple linear model for the
Archeops timelines with three main components: Galactic emission,
atmospheric emission and Gaussian instrumental noise.  In the time
ordered data the Galactic emission is weak relative to the noise.
Therefore, to improve the convergence of the algorithm we have used as
inputs, apart from the Archeops data, fake timelines of Galactic
emission extrapolated to the Archeops frequencies from the IRAS maps
using model 8 in \cite{finkbeiner}.  Finally, to reduce the noise
contribution we have restricted our analysis to the frequency range
from 0.03 to 2.5~Hz.  The main results of this analysis for the 143K03,
217K04, 353K01 and 545K01 bolometers are presented in
Fig.~\ref{smica_ozone}.  The blue and red curve are the reconstructed
atmospheric and Galactic emissions.  
We observe that the reconstructed
emission reproduces very well the two structures on the power spectrum
and also contributes to the multiples of the spin frequency.  As we
expect the atmospheric emission to vary with time we have performed
this analysis for different intervals.  As shown in
Fig.~\ref{smica_ozone_evol} the power spectrum of the atmospheric
parasitic noise does not change significantly neither in shape nor in
intensity.  From these results we have constructed a template of the
atmospheric emission which is subtracted from the data via a simple
decorrelation analysis as described in Sect.~\ref{sect:syste}.

\begin{figure*}[!ht] 
  \begin{center}
    \caption{From top to bottom: Galactic maps in antenna temperature for
the 143, 217, 353 and 545~GHz Archeops channels.
They are displayed in Galactic coordinates with the Galactic anticenter
at the center of the map. \label{galactic_maps}
}
  \end{center}
\end{figure*}

\subsubsection{Galactic maps}

The final Archeops Galactic maps, presented in
Fig.~\ref{galactic_maps}, are produced in the Healpix pixelization
scheme by simple coaddition of the previous processed timelines which
are previously band-pass filtered.  The low-pass filtering allows us to
both remove spurious high frequency noise in the data (see section
\ref{sect:syste}) and avoid aliasing on the final maps.  The
high-pass filtering keeps frequencies above 0.03~Hz and to reduce
ringing we first mask the brightest Galactic regions and fit an
irregularly sampled Fourier series truncated to the frequency of
interest.  The latter is then fully sampled and subtracted from the
data.  We produce individual maps for each of the detectors as well as
combined maps per channel using the best available bolometers.

From top to bottom, Fig.~\ref{galactic_maps} shows the combined Galactic maps
for the Archeops 143, 217, 353 and 545~GHz channels respectively. These are
the first available large angular scales maps of the sky in this frequency
range.  The maps are displayed in antenna temperature units and in Galactic
coordinates with the Galactic anticenter at the center of the map. The
Galactic plane structure, including for example the Cygnus region on the right
of the map and the Taurus region on the left, are clearly visible. Their
intensity increases globally with frequency as expected for Galactic dust
emission. At high Galactic latitudes the maps at low frequencies show no
contamination from the atmospheric emission. At 545~GHz we can observe some
atmospheric contamination. This was expected since the atmospheric signal is
stronger at high frequencies and because we have only a single bolometer
available. More detailed description and scientific analysis of these maps
will be presented in a forthcoming paper. In particular almost all (about 100)
identified point sources are Galactic and will be presented elsewhere.

\subsection{Specific processing for polarization maps}
\label{sec:polarmaps}

Whereas most of the preprocessing and noise subtraction is common to all
channels, the 353~GHz polarized channel requires additional specific
treatments.  The direction of polarization of the bolometers oriented
mechanically in the focal plane has to be checked, as well as their
polarization efficiencies.  Moreover, the map making algorithm differs from
that of temperature maps since it has to deal with non scalar quantities.
Details regarding the map making algorithm as well as the final polarized
Archeops maps at 353~GHz are given in \cite{archpolar}.  Previous to map
making the TOIs are processed as above using the MCMD-SMICA algorithm to
remove the contamination from atmospheric
emission.  \\

The reconstruction of the polarization directions of the bolometers in
the focal plane was performed during ground calibration.  For this
purpose, we built two wire grid polarizers of 10~cm diameters with the
same technology as the one used for wire chambers in high energy
physics experiments.  We used Cu/Be wires of 50~microns, spaced by
100~microns on circular steel frames which was expected to produce an
incoming radiation at more than 98\% at 850~microns (353~GHz).  We
built an alignment mechanism that could hold one or both polarizers on
top of the entrance window of the cryostat, facing directly the focal
plane.  One of the polarizers could rotate at 1.5~rpm.  A 13.4~Hz
chopper modulated the incoming radiation of a liquid Nitrogen
polystyrene box used as a 77~K black body to enable a lock-in
detection.  Once the lock-in and standard noise subtraction were
performed, the rotating polarizer induced a 1.5~rpm period sinusoidal
response for the polarized channels, from which the phase provided the
direction in the focal plane, and the offset and amplitude the
cross-polarization level.  The positions were confirmed to be nominal
(30, 120, 150, 240, 90, 0)~deg w.r.t. the scan axis up to the precision
of the method which was estimated to be 3~deg.  The cross polarization,
defined as the ratio cross-Intensity/co-Intensity was found to be
approximately 2\%.

The absolute calibration of the polarized photometers was performed in
the same way as for the other channels.  The relative calibration was
performed using the algorithm described in Sect.~\ref{sec:intercalib}.
Actually, it was originally designed for the polarized channels.  The
only caveat was the possible systematic effect induced by a polarized
Galactic component.  A two step iteration process was designed, the
first of which consisted in the removal of the strong polarized regions
before a final relative calibration on the non polarized parts of the
Galaxy.  This process ensured a relative calibration of the polarized
detectors better than 2\%.  More details can be found in
\cite{archpolar}.

\subsection{Specific processing for CMB maps.}

\begin{figure}[t]
  \begin{center}
    \caption{From top to bottom, power spectrum of the Archeops time ordered data
    before (black curve) and after (red curve) foreground removal for the
    143K03 and 217K04 bolometers respectively. For comparison, the bottom plot shows
    the power spectrum of the 545K01 bolometer. \label{foregroundremoval}}
  \end{center}
\end{figure}

\subsubsection{Foreground removal}

For reconstructing the CMB signal on the sky we need to remove from the
Archeops data the foreground contribution corresponding to the Galactic dust
emission and the atmospheric parasitic noise.  Both of them have a rising
electromagnetic spectrum with increasing frequency.  Therefore they are
significantly brighter in the Archeops high frequency channels.  The Galactic
dust emission in the millimetre and submillimetre range has a grey body
spectrum with an emissivity of the order of between 1.7 and 2 (see
\cite{finkbeiner,wmap,guilaine} for more details).  A slightly steeper index
increasing with frequency is observed for the atmospheric parasitic noise.

We have developed a decorrelation algorithm to remove the foregrounds
from the Archeops data.  As templates of the foreground emission we
have used the Archeops high frequency channel bolometers but also fake
timelines of the expected Galactic emission contribution to the
Archeops data to improve the efficiency of the algorithm.  
These fake timelines were produced in two steps.  First
we extrapolated the IRAS satellite data to the Archeops frequencies
using the model 8 of \cite{finkbeiner}.  Second the extrapolated maps
were deprojected into time ordered data following the Archeops scanning
strategy.  For the decorrelation analysis of each of the low frequency
(at 143 or 217~GHz) bolometers we used as templates the 353 and 545~GHz
time ordered data, fake Galactic timelines corresponding to those data
and an extra fake Galactic timeline corresponding to the decorrelated
bolometer.

To improve the efficiency of the decorrelation method we bandpass
filter both the Archeops data and the fake Galactic timelines in the
range 0.1 to 2~Hz where the atmospheric and Galactic emission dominate.
This can be clearly seen on the bottom plot of
Fig.~\ref{foregroundremoval} where we represent the power spectrum of
time ordered data for the bolometer 545K01.  At low frequency we
observe the Galactic and atmospheric emissions in the form of peaks at
frequencies which are multiples of the spinning frequency.  Between 1
and 1.5~Hz we can distinguish the atmospheric parasitic structure
discussed in Sect.~\ref{sec:specGalactic}.  Above 1.6~Hz the
instrumental noise dominates.  The correlation coefficients are computed
via a simple regression analysis using the bandpass filtered data.  A
linear combination of the templates previously smoothed and multiplied
by the correlation coefficients is removed from the data.

The first two upper plots of Fig.~\ref{foregroundremoval} show the
results of the decorrelation analysis for the Archeops bolometers
143K03 and 217K04 respectively.  We plot the power spectrum of time
ordered data before (black curve) and after (red curve) decorrelation.
For the 143K03 bolometer we can observe that the peaks in the spectrum
are completely removed by the decorrelation analysis but at frequencies
lower than 0.2~Hz.  The same is found for the 217K04 bolometer.
Further, we see that the atmospheric structures between 1
and 1.5~Hz are also removed.  The residual Galactic emission at
frequencies lower than 2~Hz increases dramatically with decreasing
frequency.  This seriously limits the size of the largest angular scale
for which the CMB angular power spectrum can be reconstructed using the
Archeops data.  Although the algorithm is very efficient residual
atmospheric and Galactic emission are expected in the final Archeops
CMB maps.  A more detailed discussion of these two issues is given in
\cite{tristram_cl} and \cite{patanchon_sz}.
\begin{figure}[t]
  \begin{center}

    \caption{{\it Top}: Angular power spectrum of the simulated CMB signal before (black line) and
      after data processing (blue line) for the 143~GHz. {\it Bottom}: Transfer function 
      of the data processing pipeline for 143~GHz data. \label{pw143}}
  \end{center}
\end{figure}

\subsubsection{The pipeline transfer function}

For an accurate determination of the CMB power spectrum with the
Archeops data we have to correct from the bias introduced in the signal
by the data processing.  For this we have estimated the Archeops
pipeline transfer function in multipole space. Here we consider the full
data processing but the destriping for which the transfer function was
discussed above.

For each of the Archeops bolometer at 143 and 217~GHz we have obtained
fake Archeops CMB timelines.  These were produced from the
deprojection, using the Archeops pointing solution, of the same
simulated CMB map.  These CMB timelines have been converted into
voltage units using the standard calibration coefficients for each
bolometer and then filtered out with the low pass prefilter (see
Sect.~\ref{sec:prefiltering}).  Further, we have added, to each
original bolometer time ordered data, the corresponding fake CMB
timeline and then saved them into files the same way the true ones
are.  Finally, we have passed each combined timeline through the full
Archeops data pipeline but through the destriping.  The effect of the
data processing on the CMB simulated signal can be easily obtained.
First we subtract from the combined processed timeline the
corresponding equally processed Archeops data.  Finally, we reproject
the difference into a map.  Comparing the CMB angular power spectrum of
the simulated Archeops CMB signal before and after data processing we
obtain the pipeline transfer function.

\begin{figure*}[t]
  \begin{center}
    \caption{From top to bottom, combined Archeops CMB maps for the 143 and 217~GHz channels.
             In the Galactic plane region the residual galactic emission is still visible but
             clearly disappears at high galactic latitudes where the CMB studies are performed.
    \label{cmb_maps}}
  \end{center}
\end{figure*}

The top panel of Fig.~\ref{pw143} shows the angular power spectrum for
the simulated CMB data before (black line) and after (blue line) data
processing for one of the 143~GHz Archeops bolometers.  Dividing the
one by the other we can estimate the Archeops pipeline transfer
function which is shown in the bottom panel plot.  We observe that the
changes induced in the CMB signal by the data processing pipeline are
smaller than 1\%.  Similar results are obtained for the other
bolometers at 143 and 217~GHz.  Therefore, there is no need to account
for the pipeline transfer function when estimating the CMB angular
power spectrum with Archeops (see \cite{archpaper,tristram_cl}).


\subsubsection{Archeops CMB maps}

The Archeops CMB maps were obtained by projection and bandpass
filtering of the foreground cleaned timelines using the Mirage optimal
map making code (\cite{mirage}).  The data were low pass filtered at
30~Hz to remove spurious high frequency noise and high pass filtered at
0.1~Hz to reduce the contribution from residual atmospheric and
Galactic emissions.  We have produced both individual maps for each of
the selected best bolometers at 143 and 217~GHz, and combined naturally
weighted maps for each of the low frequency channels.

Figure~\ref{cmb_maps} shows from top to bottom, the Archeops combined
CMB maps at 143 and 217~GHz.  These maps are in longitude rotated
Galactic coordinates with the anticenter at the center of the maps.  We
represent them in CMB temperature units.  Notice that close to the Galactic
center and in particular near by the Cygnus region (right of the map)
we can observe residuals from the Galactic emission.  However at high
Galactic latitude neither atmospheric nor Galactic residuals are
observed.  For CMB analysis (\cite{archpaper,tristram_cl}) we use a
Galactic mask to exclude the observed contaminated regions.  This mask
is overplotted in blue on the figure.  A detailed analysis of the
properties of these maps and how they compare to those from the WMAP
satellite is presented in a forthcoming paper (\cite{patanchon_sz}).
Other than CMB studies, these maps were used in combination with the
WMAP data (\cite{wmap}) to study statistically the Sunyaev-Zeldovich
effect in clusters of galaxies (\cite{carlos_sz}).

\section{Conclusions}
\label{sec:conclusions}
We have presented in this paper the full processing of the Archeops data, from
the raw telemetry to the final sky maps.  Despite intense preparatory work,
most methods and procedures discussed here were developed and implemented
after the acquisition of the real flight data.  This was mainly due to the
requirement of a large sky coverage in a short total integration time (about
24 hours) which imposed a large-circles like scanning strategy with small
redundancy and therefore making systematic effects difficult to handle.
Typically the data were contaminated by the large scale fluctuations of the
atmospheric emission and by the Galactic foreground emission.  Because of
these difficulties we were forced to apply different processing techniques to
the data for each of the main scientific goals, 1) estimation of the CMB
temperature anisotropies power spectrum, 2) study of the Galactic diffuse
emission and 3) estimation of the polarized submillimetre emission of the
Galaxy.  In particular, the destriping and filtering techniques previous to
projection on the sky were different for each pipeline leading to different
output maps.  A common general destriping, based on the assumption that the
structures do not have preferred directions on the sky, was applied to data in
all pipelines.  For CMB maps, the low frequency channels were further
decorrelated from a mixture of the high frequency data, which are dominated by
atmospheric and Galactic signal.  For Galactic maps, the atmospheric component
was subtracted using a component separation method on the timelines.  For the
polarization pipeline simultaneous time and frequency filtering was applied.  \\

The processing and the instrumental setup were improved between successive
flights going from the Trapani test flight to the latest Kiruna one.  For
example, after analysis of the data of the first two flights, we were able to
reduce significantly the high frequency noise excess in the data by moving the
spinning pivot motor higher up in the flight chain to dampen mechanical
vibrations.  A thermal dependency of the signal with the 10~K thermal stage
was completely removed for the last flight after a complete clean-up of the
corrugated back--to--back horns in each of the Archeops photometric pixels.
By contrast, because of the wearing off of the instrument from flight to
flight and the lack of time for a complete ground calibration, the instrument
was not launched in a fully optimal configuration in the latest Kiruna flight.
In particular, we are aware that as a consequence of the accidental landing
during the penultimate campaign, the telescope was slightly out of focus
producing optical beams larger and more elongated than expected.  To correct
for this
asymmetry new specific processing techniques were developed.  \\

The processing of the Archeops data could not be performed in a single
linear pipeline.  We needed extra pipelines to reconstruct the pointing
information, compute basic instrumental and observational parameters.
For example with respect to calibration, we had to design dedicated
pipelines for each of the method used: dipole reconstruction, Galaxy
dust emission, intercalibration, planet calibration.  Equally, we have
worked in independent pipelines to characterize the instrumental
response but also the data processing in order to correct the bias
introduced by those effects in the computation of the CMB temperature
angular power spectrum and of the polarized dust power spectra.  For
this purpose we have computed the transfer functions in multipole space for
the beam smoothing, the pipeline processing and the destriping.
Further, we have produced very specific pipelines to remove foreground
contributions on the data for each of the scientific goals.  Many
checks were performed on the subpipeline levels using simulated data.
Complete end-to-end tests were difficult to achieve because of the
complexity of the problem and the fact that the pipelines could not be
gathered in a single one.  \\

Archeops provides the first submillimetre maps of the sky with large sky
coverage, of the order of 30~\%, at sub--degree resolution and the
large-angular scales of both the CMB temperature anisotropies and the
temperature and polarization diffuse emission from Galactic dust.  Maps of the
temperature diffuse Galactic dust emission are available at the four Archeops
frequency channels, 143, 217, 353 and 545~GHz.  These are very useful maps as
intermediate resolution products between the FIRAS and the expected Planck HFI
maps (\cite{jpbcospar}).  Foreground-cleaned CMB maps were produced for the
lowest frequency channels at 143 and 217~GHz using the information on dust and
atmospheric emission provided by the high frequency ones.  These maps provided
the first simultaneous determination of the Sachs-Wolfe plateau and of the
first acoustic peak of the CMB anisotropies temperature power spectrum
(\cite{archpaper}) and more recently of the second acoustic peak
(\cite{tristram_cl}).  By combining those maps with the data from the WMAP
experiment and the 2MASS catalog of galaxies we obtained a local statistical
detection of the SZ effect in clusters (\cite{carlos_sz}).  In a forthcoming
paper (\cite{patanchon_sz}), an analysis of the level of any diffuse SZ
emission will be presented, using the Archeops and WMAP data in order to have
a broad electromagnetic spectral leverage.  Finally, $I$, $Q$ and $U$ maps of
the polarized diffuse emission of Galactic dust were constructed at 353~GHz
combining the measurements from the six polarized sensitive bolometers.  These
maps allowed us, for the first time, to characterize the polarized diffuse
emission from Galactic dust in the Galactic plane (\cite{archpolar}) and to
estimate the polarized power spectra of the diffuse Galactic dust emission at
intermediate and high Galactic latitudes (\cite{ponthieu05}).  \\

The Archeops data are the first available data which present very similar
characteristics to those of the Planck HFI instrument.  This is because the
instrumental configuration, the acquisition system and the scanning strategy
in Archeops and Planck are very similar.  There are few important differences
between these two data sets as for example the presence of an atmospheric
signal in the Archeops data which would not be at all present in the Planck
data and which is one of the most important systematics in Archeops.  However,
in many other aspects they are sufficiently similar to consider that the
techniques and methods applied for the processing of the Archeops data will be
of great use for processing the Planck HFI data.  In this sense, the
processing of the Archeops data was for us a learning process towards the
analysis of the Planck HFI data.  Actually, most of the preprocessing,
decorrelation, deconvolution from the bolometer time constant, beam pattern
reconstruction, noise spectrum estimation, destriping, calibration and power
spectrum estimation methods are currently being adapted to the Planck data
within the Planck HFI Level 2 data processing.

We encourage interested parties to contact members of the Archeops
collaboration to any specific scientific project (for example correlation with
other data sets) using the Archeops data.


\appendix
\section{Intercalibration procedure}
\label{app:intercalib}

We describe here the details of the intercalibration procedure of the Archeops bolometers
discussed in Sect.~\ref{sec:intercalib}.
 
\subsubsection{Modeling}

Let's formulate the problem as follows : $s_1(b), \ldots, s_N(b)$ are
$N$ profiles ({\em e.g.} of the Galaxy), measured by $N$ different
bolometers; $b$ stands for the Galactic latitude, and runs from 1 to
$B$; $\overline{s}_b$ is the estimated profile, which has to be
determined.

We can write the estimated profile as :
\begin{equation}
\label{eqn:model}
s_j(b) = \alpha_j\overline{s}(b)+n_j(b)
\end{equation}
where $n_j(b)$ is the noise in the bin $b$ of profile $j$ and
$\alpha_j$ the associated calibration coefficient.

If we assume Gaussian white noise, we can write the $\chi^2$ as :

\begin{equation}
  \chi^{\prime 2}=\sum_{j=1}^N\sum_{b=1}^B\frac{(s_j(b)-\alpha_j \overline{s}(b))^2}{\sigma_j(b)^2}
\end{equation}

with $\sigma_j(b)^2$ the noise variance $\langle n_j(b)^2\rangle$ (we
neglect noise correlations between detectors).

\subsubsection{Constraint}

We notice that the $\chi^2$ is invariant
under the following transformation :

\begin{equation}
\left\{
\begin{array}{rcl}
\displaystyle
s_k(b) & \longrightarrow & \beta \cdot \overline{s}_k(b) \nonumber \\
\displaystyle
\alpha_k & \longrightarrow & \frac{\alpha_k}{\beta} \nonumber
\end{array}
\right.
\label{eqn:transfo}
\end{equation}

This degree of freedom is due to the fact that we can only determine
the intercalibration coefficient up to a constant factor.  We must
choose a constraint on the parameters in order to converge to a unique
solution of the equations.

Let's choose a general relation :
$g\left(\{\alpha_i\},\{\overline{s}(b)\}\right) = 0$. Using the method
of Lagrange's multiplier, we thus have to minimize the function

\begin{eqnarray}
\chi^2\left(\{\alpha_i\},\{\overline{s}(b)\},\lambda\right) & = &
 \chi^{\prime 2}\left(\{\alpha_i\},\{\overline{s}(b)\}\right) \nonumber\\
& & {}+\lambda
g\left(\{\alpha_i\},\{\overline{s}(b)\}\right)
\end{eqnarray}

with respect to $\{\alpha_i\},\{\overline{s}(b)\}$ and $\lambda$.
The conditions of minimum lead to the three equations :

\begin{equation}
\label{eqn:minichi2lam}
\frac{\partial\chi^2}{\partial\lambda} =0 \Rightarrow
g\left(\{\alpha_i\},\{\overline{s}(b)\}\right) = 0
\end{equation}

\begin{equation}
\label{eqn:minichi2sbar}
\frac{\partial\chi^2}{\partial\overline{s}(b)} =
-2\sum_{i}\frac{\alpha_i (s_i(b) -
\alpha_i\overline{s}(b))}{\sigma_{ib}^2} + \lambda\frac{\partial
g}{\partial \overline{s}(b)}=0
\end{equation}

\begin{equation}
\label{eqn:minichi2alp}
\frac{\partial\chi^2}{\partial\alpha_i} =
-2\sum_{b}\frac{\overline{s}(b) (s_i(b) -
\alpha_i\overline{s}(b))}{\sigma_{ib}^2} + \lambda\frac{\partial
g}{\partial\alpha_i}=0
\end{equation}

Multiplying Eq.~(\ref{eqn:minichi2sbar}) by $\overline{s}(b)$ and
summing over $b$, and multiplying Eq.~(\ref{eqn:minichi2alp})
and summing over $i$, we find that $\lambda = 0$ is the only solution.
We thus find the two following relations :

\begin{equation}
  \overline{s}(b)=\frac{\sum_j\frac{\alpha_j s_j(b)}{\sigma_j(b)^2}}
{\sum_j\frac{\alpha_j^2}{\sigma_j(b)^2}}
\label{eqn:profmoy}
\end{equation}

and

\begin{equation}
  \alpha_k=\frac{\sum_b\frac{\overline{s}(b) s_k(b)}{\sigma_k(b)^2}}
{\sum_b\frac{\overline{s}(b)^2}{\sigma_k(b)^2}}
\label{eqn:alpha}
\end{equation}

whatever is the constraint.

These equations can be solved by iteration : starting from any set of
$\{\alpha_i\}$, we can calculate $\overline{s}(b)$ with
Eq.~(\ref{eqn:profmoy}), and calculate new $\{\alpha_i\}$ with
Eq.~(\ref{eqn:alpha}).  At each step of the iteration, we have to check
that the constraint is satisfied, or to impose it.  The iteration ends
when the relative differences between two successive steps is small
enough.

\subsubsection{Error matrix}

We can develop any $\chi^2$ functions around its minimum as~:

\begin{eqnarray}
\chi^2(\vec \Theta) & = & \chi^2\left(\vec\Theta^{(min)}\right) \nonumber\\
& & {}+ \frac{1}{2}\frac{\partial^2\chi^2}{\partial \Theta_i\partial\Theta_j}
\left(\Theta_i-\Theta_i^{(min)}\right)\left(\Theta_j-\Theta_j^{(min)}\right) \nonumber\\
& & {}+ \mathcal{O}(\Theta^3)
\end{eqnarray}

where $\vec \Theta$ is the vector of parameters and
$\vec\Theta^{(min)}$ is the minimum of the $\chi^2$.  The matrix
$\frac{1}{2}\frac{\partial^2\chi^2}{\partial \Theta_i\partial\Theta_j}$
is called the Fisher matrix (noted $F$ in the following) and is the
inverse of the matrix of correlation of the parameters, as can be
easily seen : calling $\Delta \vec \Theta$ the vector
$\vec\Theta-\vec\Theta^{(min)}$, the probability that the real
parameters are $\Theta$ can be written using the likelihood function
$\mathcal{L}=\exp(-\chi^2/2)$, {\it i.e.} :

\begin{equation}
\mathcal{L} \propto \mbox{e}^{-\frac{\Delta\vec\Theta^t\cdot F\cdot\Delta\vec\Theta}{2}}.
\end{equation}

Since $F$ is a positive definite matrix, it can be diagonalized, with
all its eigen values positive.  Let's call $M$ the change of frame
matrix, so that $ F = M^t D M$, where $D$ is diagonal, and
$\Delta\Theta^\prime = M\Delta\Theta$.  $M$ is orthogonal, so that
$M^{-1} = M^t$.  The correlation between two measurements will be given
by : $\langle\Delta\vec\Theta \Delta\vec\Theta^t\rangle = \langle M^t
\Delta\vec\Theta^\prime\Delta\vec\Theta^{\prime t} M \rangle = M^t
\langle \Delta\vec\Theta^\prime\Delta\vec\Theta^{\prime t}\rangle M$.
The central part is the correlation matrix in the diagonal frame : it
can be calculated directly to give $\langle
\Delta\vec\Theta^\prime\Delta\vec\Theta^{\prime t}\rangle = D^{-1}$.
We then deduce the correlation matrix in the original frame :
$\langle\Delta\vec\Theta\Delta\vec\Theta^t\rangle = M^t D^{-1} M = (M^t
DM)^{-1} = F^{-1}$.  All this calculation is made with the assumption
that the likelihood is very close to a Gaussian around its maximum.
We will see in the following that it is the case for our particular case.

In our particular case, the Fisher matrix F is dimension
$(N+B+1)\times(N+B+1)$, and we can compute it analytically for any
constraint $g$ :

\begin{equation}
\frac{1}{2}\frac{\partial^2\chi^2}{\partial\lambda^2} = 0
\end{equation}

\begin{equation}
\frac{1}{2}\frac{\partial^2\chi^2}{\partial\lambda\partial\overline{s}(b)}
= \frac{\partial g}{\partial \overline{s}(b)}
\end{equation}

\begin{equation}
\frac{1}{2}\frac{\partial^2\chi^2}{\partial\lambda\partial\alpha_i}
= \frac{\partial g}{\partial \alpha_i}
\end{equation}

\begin{equation}
\frac{1}{2}\frac{\partial^2\chi^2}{\partial \overline{s}(b)\partial\overline{s}(q)} = 
\sum_{j=1}^{N} \frac{\alpha_j^2}{\sigma_j(q)^2}\delta_{qb}
\end{equation}

\begin{equation}
\frac{1}{2}\frac{\partial^2\chi^2}{\partial \overline{s}(b)\partial\alpha_k}
= \frac{2\alpha_k\overline{s}(b)-s_k(b)}{\sigma_k(b)^2}
\end{equation}

\begin{equation}
\frac{1}{2}\frac{\partial^2\chi^2}{\partial \alpha_i\partial\alpha_j}
= \sum_{b=1}^{B} \frac{\overline{s}(b)^2}{\sigma_i(b)^2}\delta_{ij}
\end{equation}

(taking all the parameters at the minimum, including $\lambda=0$).

\begin{figure}[ht]
  \begin{center}

    \caption{Simulation profiles}
    \label{intercalibsimuprofiles}
  \end{center}
\end{figure}

\subsubsection{Robustness tests}

We have performed a bunch of tests in order to validate this
intercalibration method.  First, we have tested by Monte Carlo
simulations the reliability of the error bars computed using the Fisher
matrix.  Second, we have compared the influence of the choice of the
constraint on the intercalibration coefficients and their errors : when
comparing what is comparable, {\it i.e.} the {\it ratio} of
intercalibration coefficients, we found no difference neither in the
value nor in the error.  We have chosen to impose the constraint
$g(\{\overline{s}(b),\{\alpha_i\}\}) = \alpha_1 - 1$ ({\it i.e.} the
first profile has a relative calibration coefficient with respect to
the average profile of 1).  Finally, we have compared the iterative
minimization method (using alternatively Eqs.~\ref{eqn:profmoy}
and~\ref{eqn:alpha}) with a standard minimization program (Minuit, from
the CernLib).  Differences were below the numerical precision.

\begin{acknowledgements}
  We would like to pay tribute to the memory of Pierre Faucon who led the CNES
  team during several difficult and long campaigns. We thank the Russian
  recovery teams who worked under very harsh conditions to retrieve the
  precious instrument and data. The Swedish Kiruna Esrange base is warmly
  thanked for much help during the launch preparations. We thank the CNES balloon
  program, Programme National de Cosmologie, and the participating
  laboratories for financial support. We thank Juan R. Pardo for helping us
  with atmospheric modeling and Bruno B{\'e}zard for planet calibration.  The
  HEALPix package was used throughout the data analysis~(\cite{healpix}).
\end{acknowledgements}


\begin{thebibliography}{}

\bibitem[Amblard \& Hamilton (2004)]{amblardhamilton}
Amblard \& Hamilton, 2004, \aap, 417, 1189-1194

\bibitem[Bock et al. (1995)]{bock1995} Bock, J.J., Parikh, M.K., Fischer,
  M.L.\& Lange, E., 1995, Applied Optics, 34, 22

\bibitem[Bennett et al. (2003)]{wmap}
Bennett, C.~L., Halpern, M., Hinshaw, G. et al., 2003, \apj, 148, 1
  
\bibitem[Beno\^{\i}t \& Pujol (1994)]{cryostat}
Beno\^\i t, A. \& Pujol, S., 1994, Cryogenics, 34, 321
  
\bibitem[Beno\^{\i}t et al. (2002)]{trapani}
Beno\^\i t, A. et al. 2002, Astropart. Phys., 17, 101
  
\bibitem[Beno\^{\i}t et al. (2003a)]{archpaper}
Beno\^{\i}t A. et al. 2003, \aap, 399, No. 3, L19
  
\bibitem[Beno\^{\i}t et al. (2003b)]{archpaper_cospar}
Beno\^{\i}t A. et al. 2003, \aap, 399, No. 3, L25
  
\bibitem[Beno\^{\i}t et al. (2004)]{archpolar}
Beno{\^ i}t, A., et al.\ 2004, \aap, 424, 571 

\bibitem[Bernard (2004)]{jpbcospar}
Bernard, 2004,35th COSPAR Scientific Assembly, 4558 

\bibitem[Bourrachot (2004)]{bourrachotthesis}
Bourrachot.\ 2004, Universit\'e Paris Sud - Paris XI, 
http://tel.ccsd.cnrs.fr/documents/archives0/00/00/
77/02/index\_fr.html

\bibitem[Bock et al. (1996)]{Bock:1996} Bock, J. J, DelCastillo, H. M.,
  Turner, A. D., Beeman, J. W., Lange, A. E.  Mauskopf, P. D., 1996,
  Proc.  of the 30th ESLAB symp. on ``Submillimetre and Far-Infrared
  Space Instrumentation'', ESA-ESTEC 1996
  
\bibitem[Borrill (1999)]{madcap} Borrill, J., 1999, In {\it Proceedings of the
    5th European SGI/Cray MPP Workshop\/}, Bologna, Italy, astro-ph/9911389
  
\bibitem[Camus et al. (2000)]{Camus:2000} Camus, Ph., Bergé, L.,
  Dumoulin, L., Marnieros, S., \& Torre, J.--P. 2000, Nucl. Instr. and
  Methods in Phys. Res. A444, 419

\bibitem[Cappellini et al. (2003)]{Cappellini:2003} Cappellini, B. Maino, D.,
  Albetti, G., {\it et al. \/}, 2003, \aap, 409, 375

\bibitem[Chattopadhyay et al. (1999)]{omt}
Chattopadhyay, G., {\it et al.\/}, 1999, IEEE microwave and guided wave letters
  
\bibitem[Delabrouille et al. (2003)]{smica}
Delabrouille J.,Cardoso J.F., Patanchon,G., 2003, \mnras, 346, 1089

\bibitem[Dickinson et al. (2004)]{vsapaper}
 Dickinson, C. et al. 2004, MNRAS, 353, 732
 
\bibitem[Dor\'{e} et al. (2001)]{mapcumba} Dor\'{e}, O., Teyssier, R.,
  Bouchet, F. R., Vibert, D., \& Prunet, S., 2001, \aap, 374, 358
  
\bibitem[Dragone (1982)]{dragone}
Dragone, C. 1982, IEEE Trans. Ant. Prop., AP-30, No. 3, 331
  
\bibitem[Ferreira \& Jaffe (2000)]{ferreira}
Ferreira, P. \& Jaffe, A. H., 2000, MNRAS, 312, 89

\bibitem[Finkbeiner et al (1999)]{finkbeiner}
Finkbeiner, Douglas P., Davis, Marc, Schlegel, David J., 1999, \apj, 524, 867

\bibitem[Fixsen et al (1997)]{fixsen} Fixsen, D. J., Weiland, J. L., Brodd,
  S., Hauser, M. G., Kelsall, T., Leisawitz, D. T., Mather, J. C., Jensen, K.
  A., Schafer, R. A., Silverberg, R. F., 1997,\apj, 490, 482

\bibitem[Goldin et al. (1997)]{goldin:1997}
Goldin, A.B. {\it et al.}, 1997, \apjl, 488, L161

\bibitem[Gorski et al. (1999)]{healpix} Gorski~K.~M., Hivon~E. \&
  Wandelt~B.~D., 1999, in Proceedings of the MPA/ESO Cosmology Conference
  "Evolution of Large-Scale Structure", eds. A.J. Banday, R.S. Sheth and L. Da
  Costa, PrintPartners Ipskamp, NL, pp. 37-42 (also {\tt astro-ph/9812350},
  http://www.eso.org/science/healpix)

\bibitem[Halverson et al. (2002)]{dasi}
Halverson, N. W. et al. 2002, \apj, 545, L5

\bibitem[Hanany et al. (2000)]{maxima}
Hanany, S. et al. 2000, ApJ, 545, L5
  
\bibitem[Hern\'andez-Monteagudo et al. (2006)]{carlos_sz}
  Hern\'andez-Monteagudo, C., Mac\'{\i}as-P\'erez, J.F., Tristram, M. \&
  Desert, F.--X., 2006, \aap, 449, 41

\bibitem[Hinshaw et al. (2003)]{wmapproc}
Hinshaw, G. et al., 2003, \apj, 148, 63

\bibitem[Hivon et al. (2002)]{master} 
Hivon, E., Gorski, K.~M., Netterfield, C.~B. et al., 2002, \apj, 567, 2
  
\bibitem[Janssen \& Gulkis (1992)]{gulkis} Janssen, M. A., \& Gulkis, S.,
  Mapping the Sky with the COBE-DMR. In ``The Infrared and Submillimetre Sky
  after COBE'', M. Signore and C. Dupraz editors, Dordrecht, Kluwer, 1992

\bibitem[Kuo et al. 2004]{acbarproc}
Kuo, C.~L. et al. 2004, \apj, 600, 32 

\bibitem[Lagache (2003)]{guilaine}
Lagache, G., 2003, \aap, 405, 813L  

\bibitem[Lagache \& Douspis (2006)]{lagadou_calib} 
Lagache, G., Douspis M., \aap, submitted

\bibitem[Lamarre, J.M. et al. (2003)]{lamarreplanck}
Lamarre, J.M. et al. 2003, New Astronomy Reviews, 47, 11-12

\bibitem[Lee et al. 2001]{maxiproc2}
Lee,A.~T. et al., 2001, \apj, 516, 1L

\bibitem[Mac\'\i as--P\'erez \& Bourrachot, (2006)]{wavpaper}
Mac\'\i as--P\'erez \& Bourrachot, A., 2006, \aap, accepted

\bibitem[Masi et al. (2005)]{boomproc2}
Masi, S. et al. 2005, \apj, submitted, {\tt astro-ph/0507509}

\bibitem[Mason et al. (2003)]{cbi}
Mason, B. S. et al. 2003, \apj, 591, 540

\bibitem[Mather et al. (1999)]{Mather:1999} Mather, J. C., Fixsen, D.
  J., Shafer, R. A., \etal, 1999, ApJ, 512, 511
\bibitem[Mather et al. 1990]{Mather:1990} 
  Mather, J.~C.,Cheng, E.~S., Shafer, R.~A., Wright, E.~L., Meyer, S.~S., Weiss, R., Fixsen, D.~J., Eplee, R.~E., Isaacman, R.~B., Read, S.~M., 1990, \baas, 22, 1216

\bibitem[Mather (1984)]{Mather:1984} Mather, J. C., 1984, Applied
  Optics, 23, 584

\bibitem[Moreno (1998)]{Moreno:1998} Moreno, R., PhD Thesis, ParisVI University
  
\bibitem[Mizuguchi, Akagawa \& Yokoi (1978)]{mizu}
Mizuguchi, Y., Akagawa, M., \& Yokoi, H., 1978, Elect. Comm. in Japan, 61-B, No. 3, 58

\bibitem[Nati \etal (2003)]{Nati:2003} Nati, F., de Bernardis, P., Iacoangeli,
  A., Masi, S., Benoit, A., Yvon, D., 2003, Review of Scientific Instruments, 
  74, 4169
 
\bibitem[Netterfield et al. (2002)]{boomerang}
Netterfield, C. B. et al. 2002, \apj, 568, 38
  
\bibitem[Pardo \etal, (2002)]{Pardo:2001} 
 Pardo, J., Cernicharo, J., Serabyn, E., 2002,
 Astronomical Site Evaluation in the Visible and Radio Range. ASP Conference Proceedings, Vol. 266. 
 Edited by J. Vernin, Z. Benkhaldoun, and C. Mu\~noz-Tu\~n\'on, Astronomical Society of the Pacific, p.188

\bibitem[Piat \etal (2001)]{piat:2001} 
Piat, M., Torre, J.-P., Lamarre, J.-M., Beeman, J., Bathia, R.S., 2001, Journal of 
Low Temperature Physics, 125, 5--6

\bibitem[Piat \etal (2003)]{piat:2003} 
Piat, M., Lagache, G., Bernard, J.P., Giard, M., Puget, J.-L., 2003, \aap, 393, 359

\bibitem[Patanchon \etal (2005)]{patanchon_sz} 
Patanchon, G., Tristram, M., Mac\'{\i}as-P\'erez, J.F. {\it et al}, 2005, \aap, in preparation

\bibitem[Press et al. (1992)]{numrec}
Press, W. H., Flannery, B. P., Teukolsky, S. A. \& Vetterling, W. T., {\it Numerical Recipe}, Cambridge
  University Press, 1992
  
\bibitem[Ponthieu et al. 92005)]{ponthieu05}
Ponthieu, N., Mac\'\i as--P\'erez, J.~M., Tristram, M., et al, 2005,
\aap, 444, 327

\bibitem[Prunet et al. (2000)]{prunet}
Prunet, S., Netterfield, C. B., Hivon, E. \& Crill, B. P., In {\it Proceedings of the XXXVth Rencontres de
    Moriond, Energy densities in the Universe.\/} Edition Fronti\`{e}res, 2000, astro-ph/0006052

\bibitem[Rabii et al. 2005]{maxiproc4}
Rabii, B. et al. 2005, \apj, submitted, astro-ph/0309414

\bibitem[Ruhl et al. 2003]{boomproc1}
Ruhl, J.~E. et al. 2003, \apj, 599, 786

\bibitem[Smoot et al. (1991)]{paperdipole}
Smoot, G. F. et al. 1992, \apj, 371, L1

\bibitem[Smoot et al. (1992)]{smoot}
Smoot, G. F. et al. 1992, \apj, 395, L1

\bibitem[Stompor et al. 2002]{maxiproc3}
Stompor, R.. et al. 2002, \prd, 65, 2, 22003

\bibitem[Tristram et al. (2004)]{asymfast}
Tristram, M., Hamilton, J.-Ch., Mac\'\i as-Per\'ez, J.~F. et al., 2004, \prd, 69, 123008

\bibitem[Tristram et al. (2005a)]{xspect}
Tristram, M., Mac\'{\i}as-Per\'ez, J.~F., Renault, C. et al., 2005, \mnras, 358, 833

\bibitem[Tristram et al. (2005b)]{tristram_cl}
 Tristram, M., Patanchon, G.,  Mac\'{\i}as-P\'erez, J.F.  et al. 2005, \aap, 436, 785

\bibitem[Yvon \& Mayet (2005)]{mirage}
Yvon D. \& Mayet F., 2005, \aap, 436, 729

\end{thebibliography}
\end{document}